# Visualization of two-dimensional transition dipole moment texture in momentum space using high-harmonic generation spectroscopy

K. Uchida[1*], V. Pareek[2], K. Nagai[1], K. M. Dani[2], and K. Tanaka[1*]

[1]*Department of Physics, Graduate School of Science, Kyoto University, Kyoto, Kyoto 606-8502, Japan*

[2]*Femtosecond Spectroscopy Unit, Okinawa Institute of Science and Technology Graduate University, Okinawa 904-0495, Japan*

**ABSTRACT**

Highly nonlinear optical phenomena can provide access to properties of electronic systems which are otherwise difficult to access through conventional linear optical spectroscopies. In particular, high harmonic generation (HHG) in crystalline solids is strikingly different from that in atomic gases, and it enables us to access electronic properties such as the band structure, Berry curvature, and valence electron density. Here, we show that polarization-resolved HHG measurements with bandgap resonant excitation can be used to probe the transition dipole moment (TDM) texture in momentum space in two dimensional semiconductors. TDM is directly related to the internal structure of the electronic system and governs the optical properties. We study HHG in black phosphorus, which offers a simple two-band system. We observed a unique crystal-orientation dependence of the HHG yields and polarizations. Resonant excitation of band edge enables us to reconstruct the TDM texture related to the inter-atomic bonding structure. Our results demonstrate the potential of high harmonic spectroscopy for probing electronic wavefunctions in crystalline solids.

The transition dipole moment (TDM) is a fundamental concept describing light-matter interactions and governs the optical properties of materials. In solid systems, the TDM texture $\boldsymbol{D}_{vc}(\boldsymbol{k})$ in three-dimensional momentum space reflects the nature of the electronic states, such as atomic orbitals and inter-atomic bonding, and provides crucial insight into the internal structure of solids, such as the band structure. Polarization-resolved absorption measurements enable us to access the TDM at the band edge, where the optical response is mainly attributed to a single point in momentum space. However, it is usually difficult to access its texture far away from the band edge. This is because the contributions of the TDM to the conventional linear optical response, such as absorption, are averaged over the isoenergy lines.

Recently, high harmonic generation (HHG) in solids [1-14], which is attributed to Bloch electron dynamics under a strong laser field, has been getting more attention as a tool for probing the electronic structures as in the study of a molecular system [15]. The HHG process in solids can be described as a three-step model analogous to that of gaseous media [16]: creation, acceleration, and recombination of electron-hole pairs. Acceleration of electron-hole pairs in each band induces so-called intraband current, which carries information about the band structure. By using this characteristic of intraband current, several previous reports have tried to obtain the band dispersion along the applied electric field direction [3]. On the other hand, the creation and recombination of electron-hole pairs, which induces HHG, originates from interband polarization and is governed by TDM $\boldsymbol{D}_{vc}(\boldsymbol{k})$. Several experiments on HHG in solids reported that HHG yields above the bandgap energy are strongly linked to the joint density of states (JDOS), and interband HHG dominates intraband HHG above bandgap energy [11,12,17,18]. This result suggests that interband HHG can be utilized as a probe of TDM texture [19,20]. So far, HHG measurements have been performed by using intense light far off-resonant with the bandgap energy of the target sample. In this limit, a number of electronic bands in entire Brillouin zone involves in HHG process, and it is difficult to disentangle the relation between HHG properties and TDM texture in momentum space. In contrast, the tuning of the driving frequency to a certain transition gives us an opportunity to select a pair of electronic bands and create electron-hole pairs with a defined crystal momentum [21].

In this report, we measured polarization-resolved HHG spectra under the resonant excitation at the band edge (interband-resonant HHG) in the thin-layer bulk black phosphorus (BP). Using HHG data, we succeeded in the reconstruction of TDM texture in the two-dimensional momentum space. This means that the interband-resonant HHG provides an efficient tool to probe the TDM texture.

Figure 1 (a) shows a schematic diagram of the interband-resonant HHG process in the two-band system. In the conventional HHG process, the driving laser field is usually far below resonance with the bandgap energy ($\varepsilon_g(\boldsymbol{k}_i)$, where $\boldsymbol{k}_i$ is the crystal momentum of the band edge), resulting in electron-hole creation through tunneling processes in the entire Brillouin zone [22]. On the other hand, when the driving laser field is resonant with the bandgap energy, coherent electron-hole pairs are selectively created at the band edge $\boldsymbol{k}_i$. This condition is similar to that in high-order sideband generation, which selectively prepares coherent electron-hole pairs at the band edge by injecting an additional NIR pulse [22-26]. The created electron-hole pairs are simultaneously accelerated along the driving field direction and then recombine. As shown in Sec. 8 of the Supplemental Information [27], when the contribution of the interband polarization in the HHG process is much stronger than that of the intraband current, the polarization and amplitude of the HHG emission are determined by the TDM at the crystal momentum $\boldsymbol{k}_r$ where recombination occurs. One can directly derive an approximate TDM $\boldsymbol{D}_{cv}^*$ for a system with inversion symmetry and a simple band structure through the interband polarization current $\boldsymbol{j}_{er}(\omega)$ as follows[27]:

$$\boldsymbol{D}_{cv}^*(\boldsymbol{k}_r) \propto \boldsymbol{j}_{er}(\omega)/(\alpha(\theta, F_0, \omega, \Omega)\boldsymbol{D}_{cv}(\boldsymbol{k}_i) \cdot \boldsymbol{F}_0) \quad . \tag{1}$$

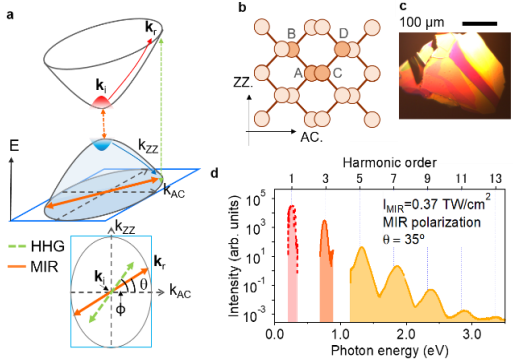

FIG 1. (a) Schematic diagram of interband-resonant HHG process in crystal momentum space. Orange solid arrows indicate driving electric field direction. The upper figure shows that electron-hole pairs are created at $\boldsymbol{k}_\mathrm{i}$ (orange dashed line) and then accelerated in the direction of the driving field in their respective bands (red and blue solid lines), and they finally emit radiation at $\boldsymbol{k}_\mathrm{r}$ (yellow green dashed line). $\boldsymbol{k}_\mathrm{AC}$ and $\boldsymbol{k}_\mathrm{ZZ}$ respectively denote the crystal momentum components along the armchair (AC) and zigzag (ZZ) directions of BP. The lower panel shows the same diagram projected on a two-dimensional plane. Here, θ(ϕ) is defined as the angle between the driving field (transmission axis of the polarizer for HHG emission) and the AC direction. The gray solid line represents the isoenergy line. (b) Crystal structure of puckered layer black phosphorus from the upper side. A, B, C, and D indicate four sublattices. (c) Image of the black phosphorus flake used in the experiment. The horizontal direction corresponds to the AC direction. (d) Typical HHG spectrum from black phosphorus with MIR intensity $I_\mathrm{MIR}$ of 0.37 TW/cm$^2$ at focal point in air and MIR polarization θ = 35°.

Here, $\hbar\Omega$ and $\hbar\omega$ are respectively photon energies of incident driving field and high harmonic emission, $\boldsymbol{F}_0 = F_0(\cos\theta, \sin\theta)$ describes the direction of the driving electric field, and $\boldsymbol{k}_\mathrm{r}$ satisfies the relations $\varepsilon_g(\boldsymbol{k}_\mathrm{r}) = \hbar\omega$ and $(\boldsymbol{k}_\mathrm{r} - \boldsymbol{k}_\mathrm{i}) \parallel \boldsymbol{F}_0$. $\alpha(\theta, F_0, \omega, \Omega)$ contains the information about electron-hole dynamics and depends on $\boldsymbol{k}_\mathrm{r}$. In a simple band structure, a set of $\boldsymbol{k}_\mathrm{r}$ form a single closed isoenergy line in momentum space, and $\alpha$ can be assumed to be almost constant with respective to the direction of $\boldsymbol{k}_\mathrm{r}$, i.e., MIR polarization angle θ [28]. Since the interband polarization current $\boldsymbol{j}_{er}(\omega)$ induces the HHG emission, the TDM $\boldsymbol{D}^*_{cv}(\boldsymbol{k}_\mathrm{r})$ is parallel to the polarization of the emission, and its amplitude is scaled by the efficiency of the initial electron-hole pair creation represented by the inner product of $\boldsymbol{D}_{cv}(\boldsymbol{k}_i)$ and $\boldsymbol{F}_0$. Therefore, we can extract an approximate TDM texture $\boldsymbol{D}^*_{cv}(\boldsymbol{k}_\mathrm{r})$ on the isoenergy lines at the HHG emission energies in solids through measurement of the crystal orientation dependence of the HHG yield and polarization when $\boldsymbol{D}_{cv}(\boldsymbol{k}_i)$ is known and the interband polarization is dominant in the HHG process.

To demonstrate probing of TDM texture, we chose thin-layer bulk BP as a sample. Bulk BP is a direct bandgap semiconductor with a bandgap energy of 0.3 eV located at the Z point [29]. Figure 1(b) shows the top view of the crystal structure of puckered layer BP, which contains four atoms (A, B, C, and D) in its primitive cell. Owing to its puckered honeycomb lattice structure belonging to the D$_{2h}$ point group, the electronic system of BP can be regarded as a two-band system with an effective mass much lighter along the armchair(AC) direction than along the zigzag(ZZ) direction for both electrons and holes [30,31]. This two-band system without degeneracy, except for the spin degree of freedom, gives TDM at the band edge $\boldsymbol{D}_{cv}(\boldsymbol{k}_i)$ parallel to the AC direction and causes linear dichroism [30-33]. In accordance with the effective mass and TDM near the band edge, the linear and perturbative-nonlinear optical properties can be usually regarded as those of a quasi-one-dimensional system along the AC direction [32-35]. On the other hand, although the optical absorption between two bands averaged over the crystal momentum is always stronger for AC-polarized light up to 3 eV [32,33], the TDM texture far away from the band edge region is nontrivial. A simulation of HHG suggests that the interband polarization dominates the intraband current in bulk BP [36]. Therefore, this simple two-band system is an ideal platform with which to investigate the interband-resonant HHG process discussed above.

We prepared thin-layer BP samples on a fused silica substrate by using the scotch-tape-exfoliation method, as shown in Fig. 1(c). The layer thickness of the samples was estimated to be ~30 nm, which is thick enough for the sample to be regarded as bulk [37]. Since the layer thickness is much thinner than the wavelength of the incident light and the HHG emission, we can neglect propagation effects inside the sample, which modify the intrinsic HHG signal [38-40]. The crystal orientation of the sample was determined by using linear dichroism [41]. We used 60 fs MIR pulses at a photon energy of 0.26 eV as the driving field of the electronic system. The high energy region of MIR spectrum covers bandgap energy of black phosphorus (0.3 eV), indicating that our experimental condition can be regarded as that of the interband-resonant HHG process [42]. We confirmed that the effect of liner dichroism on incident MIR electric field strength is negligible [43]. The angle between the linearly polarized MIR field and the AC direction is denoted as θ, and the polarization of the HHG emissions was resolved by a wire-grid polarizer (lower panel of Fig. 1(a)).

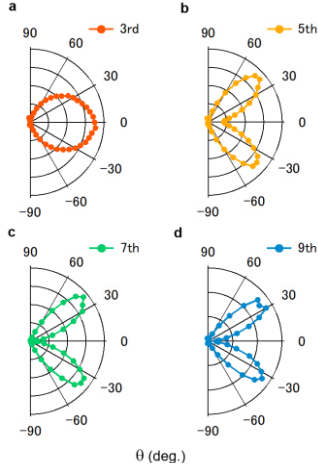

FIG 2. Crystal orientation dependence of HHG yields for the (a) third, (b) fifth, (c) seventh, and (d) ninth harmonics.

Figure 1(d) shows a typical HHG spectrum from the sample with a layer thickness of 30 nm. We succeeded in measuring from third to thirteenth order harmonics. We could not observe any even-order harmonics since BP has inversion symmetry. The HHG yield shows a gradual decrease with increasing harmonic order without any resonant feature. The gradual decrease supports the conclusion that BP can be regarded as a simple two-band system without a van-Hove singularity for the optical response in the visible range since a van-Hove singularity in the JDOS causes an enhancement of the HHG emission [11,12,17].

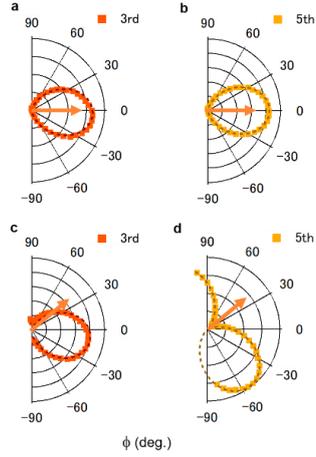

FIG 3. (a, b) Polarization state of the (a) third and (b) fifth harmonics for MIR polarization $\theta = 0°$. (c, d) Polarization state of the (c) third and (d) fifth harmonics for MIR polarization $\theta = 40°$.

To access the anisotropic HHG response, which strongly depends on the crystal structure and electronic states [5,7,8,10,14], we measured the crystal orientation dependence of the HHG yield, as shown in Figs. 2(a)-(d). For all the harmonics, the HHG yield is strongly suppressed along the ZZ direction ($\theta = 90°$). This is in good agreement with the part of Eq. (1), $|\boldsymbol{D}_{cv}(\boldsymbol{k}_i) \cdot \boldsymbol{F}_0|^2 \propto \cos^2\theta$, which describes the creation of coherent electron-hole pairs and becomes zero for $\theta = 90°$. According to Eq. (1), the deviation of the crystal orientation dependence from $\cos^2\theta$ reflects the anisotropy of the TDM amplitude. For the third-order harmonics (Fig. 2(a)), the crystal orientation dependence is similar to $\cos^2\theta$, indicating almost uniform TDM amplitudes. In contrast, HHG yields are suppressed along the AC direction ($\theta = 0°$) and reach a maximum for $\theta \sim 40°$. This tendency becomes stronger as the harmonic order increases. These results imply that the TDM amplitude is relatively small for $\boldsymbol{k}_r$ along the AC direction in the region far away from the band edge $\boldsymbol{k}_i$. Hence, we cannot apply the quasi-one-dimensional picture to the highly nonlinear optical process in BP, in contrast to its linear optical and transport responses.

We also checked the polarization states of the HHG emission depending on the crystal orientation, which reflects the direction of the TDM. Figures 3(a) and (c) show the polarization of the third-order harmonics for driving fields along θ = 0° and 40°. The third-order harmonics are polarized almost along the AC direction independently of the incident MIR polarization. This is consistent with the quasi-one-dimensional nature of BP near the band edge. In contrast, the polarization of the fifth-order harmonics highly depends on the incident MIR polarization. As shown in Fig. 3 (b), for θ = 0°, the fifth-order harmonics are linearly polarized along the AC direction, like the third harmonics (Fig. 3(c)). On the other hand, for θ = 40°, the polarization is not along the AC direction and is almost perpendicular to the incident MIR polarization (Fig. 3(d)). Such an almost perpendicular polarization of odd-order harmonics has been never observed in previous reports [7,8,10]. There are also perpendicularly polarized HHG signals for the higher harmonics, and this suggests that $D_{cv}(k_r)$ is almost perpendicular to $k_r$ for θ = 40° in the region far away from the band edge $k_i$ in accordance with the interband-resonant HHG process [44]. Note that this unique behavior cannot be obtained from the intraband current based on the band structure of bulk BP, supporting that the dominant contribution should be the interband-resonant HHG process [45]. These results indicate that not only the HHG yield, but also the HHG polarization shows a response strongly deviated from that of a quasi-one-dimensional system.

On the basis of the experimental results and the assumption that the interband-resonant HHG process is dominant, we can reconstruct the approximate TDM texture $D_{cv}(k)$ on the isoenergy lines at the third (0.8 eV), fifth (1.3 eV), seventh (1.8 eV), and ninth (2.3 eV) harmonics emission energies by using Eq. (1) [46]. Figure 4 (a) shows the experimentally obtained TDM texture on isoenergy lines calculated with the tight-binding model [47,48]. The following are characteristic features of the obtained TDM structure in bulk BP: (1) the TDM amplitude is relatively weak for $k$ along the AC axis (θ = 0°), (2) As θ increases from 0°, $D_{cv}(k)$ turns in the direction opposite to $k$, and its component perpendicular to $k$ becomes large, (3) The above tendencies becomes apparent as $|k - k_i|$ increases.

We also calculated the TDM texture based on the tight-binding model by considering nearest-neighbor interatomic bonding between the A(C) and B(D) atoms (Fig. 1(b)). As shown in Fig. 4(b), the resultant TDM texture on the isoenergy lines qualitatively agree with characteristics (1) to (3) in the experimentally obtained one. This suggests that the unique TMD texture in bulk BP far away from the band edge can be attributed to the inter-atomic bonding structure of BP and that it determines HHG properties deviated from that of a quasi-one-dimensional system. Note that there is a quantitative gap between the experiment and calculation based on the tight-binding model. The experimentally obtained TDM direction is more tilted towards the ZZ direction compared with the calculated one. Also, The TDM amplitude is more strongly suppressed in the AC direction in the experiment than in the calculation. These differences indicate that the ZZ component of the TDM is several times relatively larger in the experiment than in the calculation. This may be attributed to differences in the details of their electronic wavefunctions and contributions from distant inter-atomic bonds. Figures 4 (c-f) show the HHG properties calculated with the corrected TDM texture in which the ZZ component is three times larger than that of the tight-binding model (Fig. 4(b)), so that the TDM texture quantitatively agrees with the experimental one. The corrected TDM texture is provided in the Supplementary Information (Fig. S14(c)) [49]. The calculated crystal orientation dependence of the HHG yield and polarization both match those of the experiment. These results support the validity of the interband-resonant HHG process discussed above.

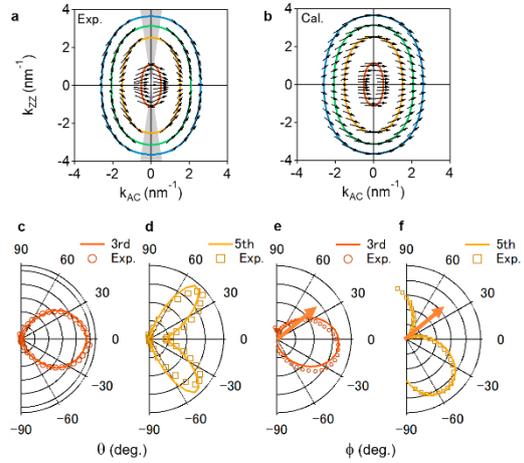

FIG 4. (a) Experimentally obtained in-plane TDM texture. The origin is set to be the Z point. Red, orange, green, and blue solid lines show isoenergy lines corresponding to the emission energy of the 3rd (0.8 eV), 5th (1.3 eV), 7th (1.8 eV), and 9th (2.3 eV) harmonics, respectively. The peak amplitude of the TDM vector on each isoenergy line is normalized for clarity. The gray shaded area indicates the region where we cannot obtain TDM due to the small signal-to-noise ratio. (b) Calculated TDM texture corresponding to (a) without correction. (c,d) Normalized crystal orientation dependence of the third (c) and fifth (d) harmonics yields. (e,f) Normalized polarization states of (e) the third and (f) fifth harmonics for θ = 40°. Solid lines in (c-f) indicate the ones calculated by using the corrected TDM texture in which the ZZ component is three times that of the tight-binding model (b). The corrected TDM is provided in the Supplementary information (Fig. S14(c)) [49]. Open circles and squares in (c-f) indicate the corresponding experimentally obtained data (same as Fig. 2 (a,b) and Fig. 3 (c,d)).

In conclusion, we experimentally visualized TDM texture in two-dimensional momentum space for thin-layer bulk black phosphorus (BP). We observed HHG from the BP under the resonant condition, which can be regarded as a simple two-band system, from the infrared to visible range. We used these experimental results to reconstruct the TDM texture on the isoenergy lines at the HHG emission energies. The obtained texture qualitatively matched the calculation considering interatomic bonding, indicating that the TDM texture far away from the band edge reflects the internal structure of crystalline solids. Here, we observed unique HHG signals directly linked to the TDM texture in a simple two-band system in BP. On the other hand, the relation between the TDM texture and the resultant HHG signal is more complicated and nontrivial in a multi-band system with degeneracy. The limitations of our technique may be overcome by using high-order sideband generation geometry with a few-cycle CEP-stable driving field pulse. Precise control of creation and acceleration of electron-hole pairs will enable us to access the TDM texture in a more complicated system without inversion symmetry and may lead to a reconstruction of an electronic wavefunction in a solid system [24].

**ACKNOWLEDGEMENT**

This work was supported by a Grant-in-Aid for Scientific Research (S) (Grant No. 17H06124) and a JST ACCEL Grant (No. JPMJMI17F2). This work was also supported by a Kick-start fund: KICKS, Okinawa Institute of Science and Technology Graduate University. K. U. is thankful for a Grant-in-Aid for Research Activity Start-up (Grant No. 18H05850) and a Grant-in-Aid for Young Scientists (Grant No. 19K14632).

K. U., K. M. D., and K. T. conceived the research project. V. P. and K. M. D. prepared the samples. K. U. and K. N set up the experimental system. K. U., V. P., and K. N. carried out the experiments. K. U., K. N., and K. T. constructed the model and performed the calculation. K. U. and K. T. mainly prepared the manuscript. All the authors contributed to the discussion and interpretation of the result.

# Supplemental Information:
# Visualization of two-dimensional transition dipole moment texture in momentum space using high-harmonic generation spectroscopy

K. Uchida[1], V. Pareek[2], K. Nagai[1], K. M. Dani[2], and K. Tanaka[1]

[1]*Department of Physics, Graduate School of Science, Kyoto University, Kyoto, Kyoto 606-8502, Japan*

[2]*Femtosecond Spectroscopy Unit, Okinawa Institute of Science and Technology Graduate University, Okinawa, 1919-1, Japan*

## 1. Experimental setup

Figure S1(a) shows the schematic of an experimental setup. We used Ti: Sapphire regenerative amplifier (pulse width: 35 fs, pulse energy: 7 mJ, center wavelength: 800nm, repetition rate: 1 kHz) as a laser source. The part of laser output (~1 mJ) is used for generating the MIR driving field. We first generated the signal and idler outputs by using an optical parametric amplifier system (Light Conversion TOPAS-C), and then generated MIR pulses (center wavelength: 4.8 μm) by difference frequency mixing of them in a $AgGaS_2$ crystal. The transmitted signal and idler outputs are blocked by using a longpass filter (cutoff wavelength: 4 μm). Figure S1(b) shows the spectrum of driving MIR pulse. The spectrum covers bandgap energy of black phosphorus (0.3 eV), indicating the resonant excitation condition of band edge. The dip around 0.29 eV is due to $CO_2$ absorption, whose effect on the temporal profile of MIR pulse is negligible in the aspect of HHG measurement. Polarization angle and intensity of MIR pulses are controlled by two wire-grid polarizers (Thorlabs WP25M-IRA) and a liquid crystal variable retarder (Thorlabs LCC1113-MIR). MIR pulses were focused onto the sample with normal incidence by using an ZnSe lens (focal length: 62.5 mm). The spot size at the focal point was estimated to be 60 μm in full-width half-maxima by using knife edge measurement, and the pulse width is estimated to be 60 fs by using sideband measurement in $MoS_2$ as in Ref. 1. The samples were set into a vacuum container with optical windows (front: $CaF_2$, back: $BaF_2$) to avoid the degradation due to water vapor, and its position was controlled by mechanical stages so that it located at the center of MIR spot. The transmitted high harmonic emissions were corrected by UV-fused silica lens (focal length: 50 mm), and their polarization was resolved by using a wire-grid polarizer (Thorlabs WP25M-UB). The emissions were spectrally resolved by a spectrometer, and detected by a InGaAs line detector (for the third harmonics) or a Si CCD camera (for the higher harmonics).

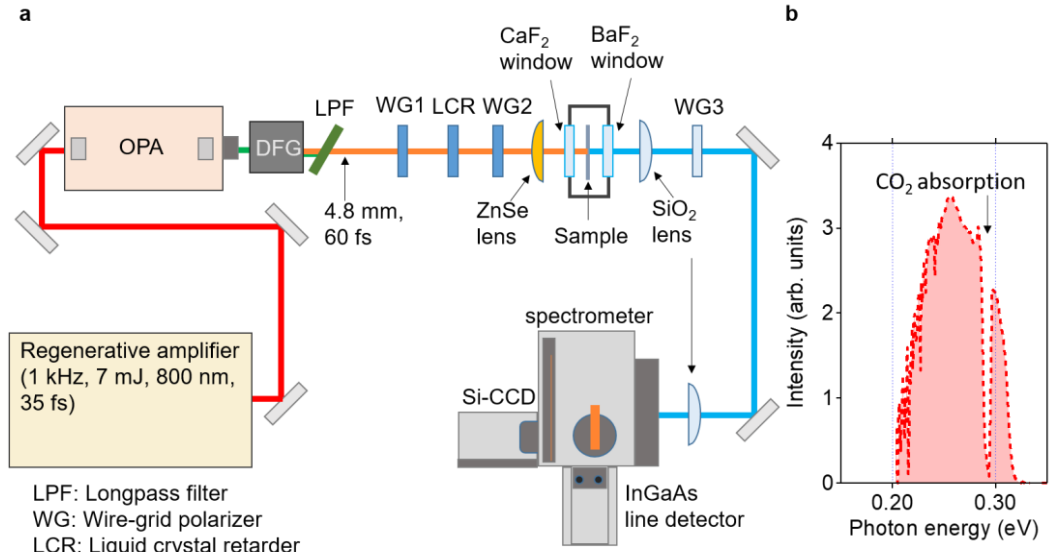

FIG S1: (a) Schematic of experimental setup. (b) Spectral profile of MIR pulse.

2. **Sample preparation**

High purity black phosphorus (BP) crystals (2D semiconductors) were used to prepare bulk BP samples on fused silica substrate. BP crystals were exfoliated onto PDMS stamps using the scotch tape method. Relatively large (> 30μm) and uniform bulk BP samples were identified using optical microscope and transferred onto fused silica substrate using a home built dry transfer setup. Optical contrast was used to approximate the thickness of the sample [2]. All the samples were prepared inside a nitrogen filled glovebox to prevent degradation of the samples. The samples were kept in inert environment (vacuum or $N_2$) for subsequent experiments and transport.

3. **Crystal orientation of bulk BP**

In bulk BP, the absorption coefficient is larger when the incident light polarization is along armchair (AC) direction than along zigzag (ZZ) direction from mid-infrared to visible region (< 3 eV). This anisotropy of absorption is called linear dichroism and can be utilized as a probe of the crystal orientation of the samples. We performed the polarization-resolved transmission measurement of bulk BP by using continuous-wave laser at 670 nm. The incident laser polarization angle was controlled by the liquid crystal retarder and the polarizer, and transmitted laser intensities were detected by the Si CCD camera. We calculated optical density ($O.D.$) given by

$$O.D.(\theta) = -\log(I_{\text{sam}}(\theta)/I_{\text{ref}}(\theta)), \tag{S1}$$

where $I_{\text{sam}}$ and $I_{\text{ref}}$ are transmitted intensity with and without sample, respectively. Figure S2(a) and (b) show respectively the sample image and the corresponding $O.D.$ as a function of incident laser polarization angle. We performed the fitting of experimental data with $A\cos^2(\theta - \theta_{AC}) + B$, and set $\theta_{AC}$ as the AC direction ($\theta_{AC} = 0°$).

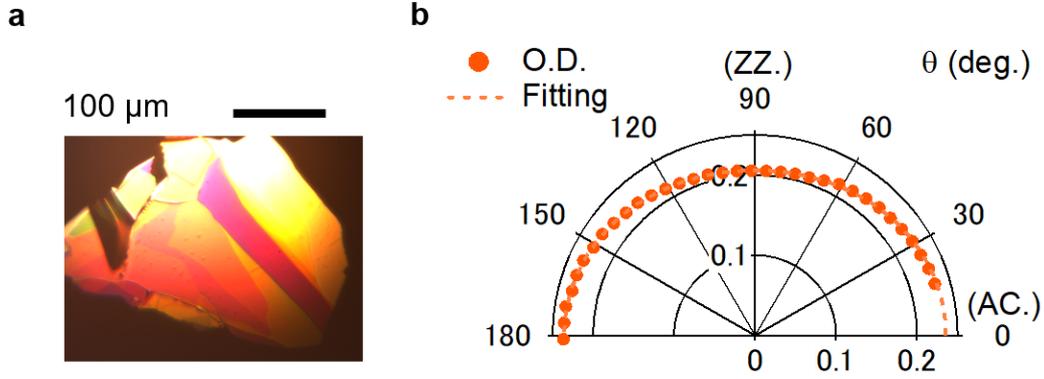

FIG S2: (a) Microscopic image of the thin layer BP sample. (b) Solid circles show obtained optical density of the sample as a function of incident laser polarization angle. Dashed line shows the fitting result.

## 4. MIR electric field inside sample

As shown in Sec. 3, BP has linear dichroism. This may cause the difference of incident MIR electric field inside the sample. To check anisotropy of incident MIR electric field, we evaluated the MIR electric field inside the sample by using a transfer matrix method with refractive indices from the literature. Figure S3(a) shows the schematic of experimental configuration. The BP flakes are attached on a fused silica substrate. MIR pulses enter from vacuum to BP so that linear and nonlinear propagation effects inside the fused silica substrate does not modify the temporal profile of MIR pulses incident to the sample. Figure S3(b) shows the estimated electric field amplitude inside the sample as a function of layer thickness. Here, we used the complex refractive indices of BP for the MIR electric field along AC direction $n_{AC}=3.16+0.0623i$, along ZZ direction $n_{ZZ}=2.83$ from the literature [3], and refractive index of fused silica $n_S=1.35$. When the layer thickness is 30 nm, the ratio of the incident MIR electric field to that inside the sample is almost isotropic, and is given by approximately 0.85, which is almost the same as the Fresnel loss at the interface between vacuum and the fused silica substrate. Hence, we neglect the anisotropy of the loss at the sample surface and estimate the electric field inside the sample by using the factor of 0.85 compared with that estimated in air. Note that MIR intensities described in the main text and the supplemental material are those estimated in air.

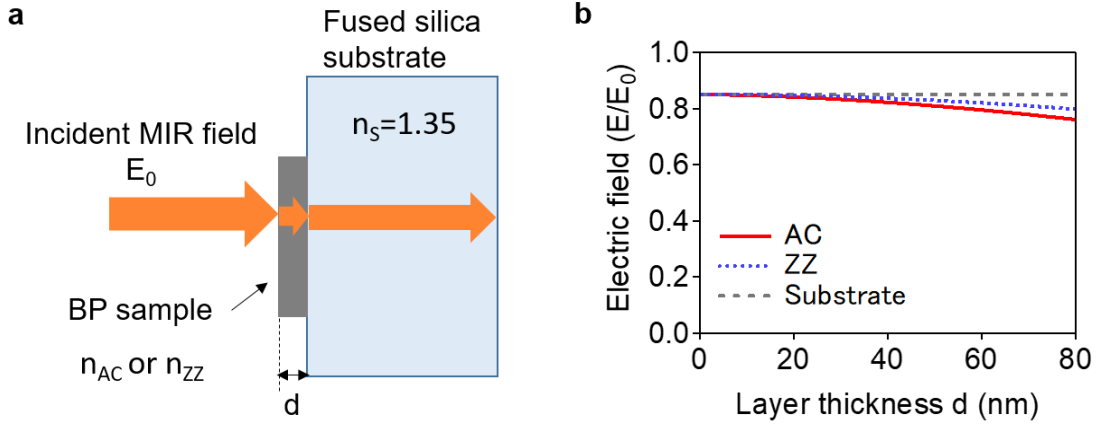

FIG S3: (a)The schematic of experimental configuration. (b) Calculated electric field inside the sample. Red solid and blue dotted lines respectively show the results using complex refractive index along AC direction and ZZ direction. Grey dashed line shows the result without the sample.

5. **Additional experimental data of HHG in BP**

Figures S4 show the polarization states of HHG in BP for $\theta = 0°$ and $40°$ corresponding to Figs. 3 in the main text. In the fifth and higher harmonics, the emissions are almost perpendicularly polarized to the incident MIR polarization for $\theta = 40°$.

We also measured the MIR intensity dependence of HHG in bulk BP. Figures S5(a) and (b) respectively show the HHG intensity as a function of the incident MIR intensity for the MIR polarization angle $\theta = 0°$ and $35°$. Above 0.1 TW/cm², HHG yields for both $\theta = 0°$ and $35°$ deviate from power law ($I_n^{HHG} \propto I_{MIR}^n$, where n is the order of high harmonics) depicted as dashed lines in Figs. S5(a) and (b). This clearly indicates the non-perturbative nonlinearity of HHG process in bulk BP above the MIR

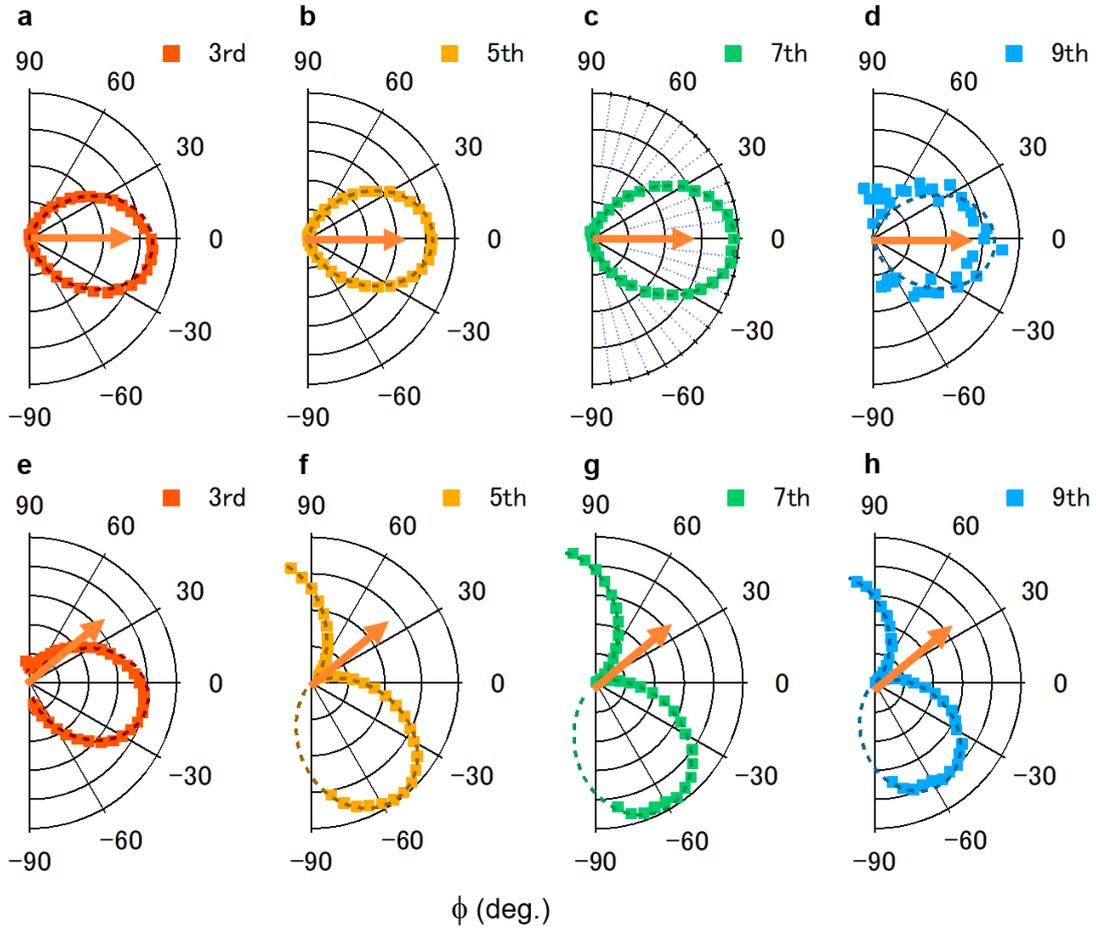

FIG S4: (a-d) Polarization state of (a) the third, (b) fifth, (c) seventh, and (d) ninth harmonics for MIR polarization $\theta = 0°$. (e-h) Polarization state of (e) the third, (f) fifth, (g) seventh, and (h) ninth harmonics for MIR polarization $\theta = 40°$

intensity of 0.1 TW/cm$^2$. In Fig. S5(b), the polarization of HHG emission is resolved. Both AC and ZZ components show similar behavior depending on MIR intensity.

Since the crystal orientation dependence is independent of MIR intensity in Eq. (1) in the main text, we also checked whether anisotropic HHG response depends on MIR intensity. Figures S6 (a)-(d) shows the polarization-resolved HHG intensities as a function of incident MIR polarization angle $\theta$ with the MIR intensities of 0.2 TW/cm$^2$ (solid line, MIR electric field strength of 10 MV/cm), 0.1 TW/cm$^2$ (dashed line, 7 MV/cm), and 0.05 TW/cm$^2$ (dashed-dotted line, 5 MV/cm). The characteristics of crystal orientation dependence are qualitatively the same for all MIR intensity, suggesting the validity of Eq. (1) in the main text. Note that the MIR polarization angle where HHG yield show maximum value slightly increase with the increase of MIR intensity. This implies several additional effects on the crystal orientation dependence which is neglected for the derivation of Eq. (1), such as the interference of possible trajectories, quantum tunneling effect. Here, we use the crystal orientation dependence with the maximum MIR intensity (0.2 TW/cm$^2$) for the reconstruction of transition dipole momentum (TDM) texture since the signal to noise ratio is the best.

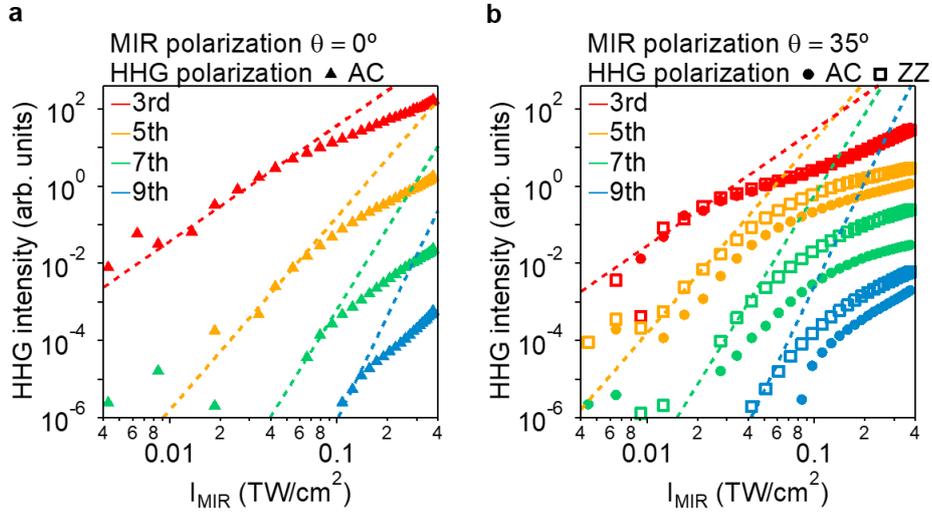

FIG S5: MIR intensity dependence of HHG yields. (a) HHG yields for $\theta = 0°$. Red, yellow, green, and blue solid triangles indicate the third, fifth, seventh, and ninth harmonic intensity. Dashed lines are proportional to $I_{\text{MIR}}^n$, which is valid in perturbative nonlinear optics. (b) HHG yield for $\theta = 35°$. Solid circles indicate HHG intensities polarized along AC direction. Open squares indicate those along ZZ direction.

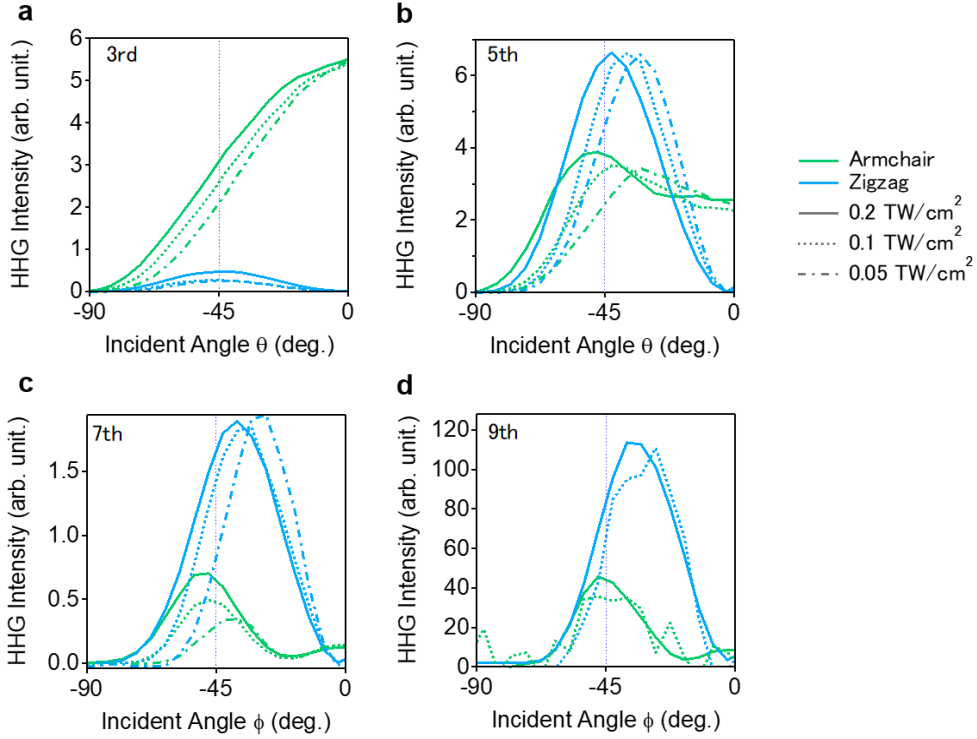

FIG S6: (a-e) Crystal orientation dependence of polarization-resolved (a) third (b) fifth, (c) seventh, (d) ninth harmonics intensity. Green(blue) lines indicate the component polarized along AC(ZZ) direction. Solid, dotted, and dashed-dotted lines respectively indicate the data with MIR intensities of 0.05, 0.1, and 0.2 TW/cm$^2$. The vertical axes for 0.1 TW/cm$^2$ and 0.05 TW/cm$^2$ are magnified for clarity.

## 6. Tight binding calculation

According to Takao et al. [4], we calculate the band structure of bulk BP. Besides, we calculate the transition dipole moment texture of bulk BP by using the eigenfunctions of bands. We consider 3s and 3p ($x$, $y$, and $z$) orbitals at atomic sites A, B, C, and D as depicted in Fig. S7. To calculate the band structure $\varepsilon_n(\mathbf{k})$ (n: band index), we solved the 16 × 16 secular equation as follows:

$$\det|H(\mathbf{k}) - \varepsilon_n(\mathbf{k})S(\mathbf{k})| = 0. \tag{S2}$$

The matrix elements of a tight-binding Hamiltonian $H(\mathbf{k})$ and a overlap integral $S(\mathbf{k})$ are given by

$$H_{i\mu j\nu}(\mathbf{k}) = \langle i,\mu,\mathbf{k}|\hat{H}|j,\nu,\mathbf{k}\rangle \quad i,j = A, B, C, D, \ \mu,\nu = s, x, y, z, \tag{S3}$$

$$S_{i\mu j\nu}(\mathbf{k}) = \langle i,\mu,\mathbf{k}|j,\nu,\mathbf{k}\rangle, \tag{S4}$$

$$|i,\mu,\mathbf{k}\rangle = \frac{1}{\sqrt{N}}\sum_{\mathbf{R}_i} \exp(i\mathbf{k}\cdot\mathbf{R}_i)|\mu,\mathbf{R}_i\rangle. \tag{S5}$$

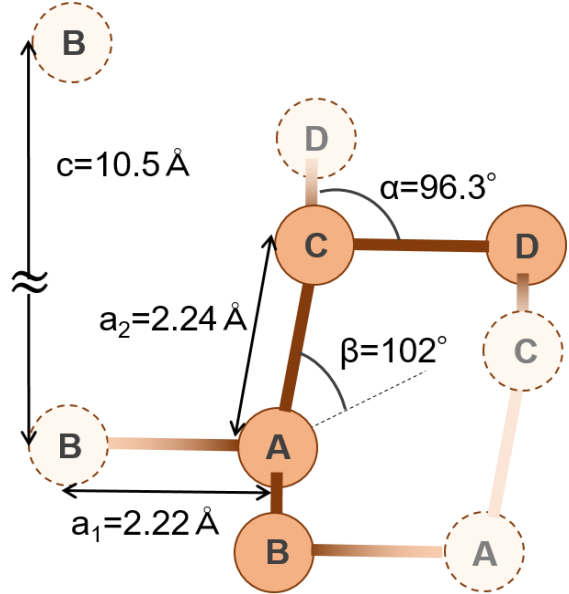

FIG S7: Schematic of the crystal structure of black phosphorus [4].

Here, $|i,\mu,\mathbf{k}\rangle$ is the basis for the tight-binding method labeled by atomic site $i$, orbital $\mu$, and crystal momentum $\mathbf{k}$. For the evaluation of $S_{i\mu j\nu}(\mathbf{k})$, we use Slater-type orbitals as follows:

$$\langle r|s, \boldsymbol{R}_i\rangle = \frac{2\xi_{3s}^{7/2}}{3\sqrt{10\pi}} |\boldsymbol{r} - \boldsymbol{R}_i|^2 \exp(-\xi_{3s}|\boldsymbol{r} - \boldsymbol{R}_i|), \tag{S6}$$

$$\langle r|x_i, \boldsymbol{R}_i\rangle = \frac{\sqrt{2}\xi_{3p}^{7/2}}{\sqrt{15\pi}} x_i |\boldsymbol{r} - \boldsymbol{R}_i|^2 \exp(-\xi_{3p}|\boldsymbol{r} - \boldsymbol{R}_i|) \; x_i = x, y, z. \tag{S7}$$

Here, we used Clementi exponents $\xi_{3s} = 1.8806 a_0$ and $\xi_{3p} = 1.6288 a_0$ from the literature [5], where $a_0$ is the Bohr radius.

To evaluate the matrix element of Hamiltonian, we used extended Huckel approximation as follows:

$$\langle \mu, \boldsymbol{R}_i|\hat{H}|\nu, \boldsymbol{R}_j\rangle = \frac{1}{2} K_{\mu\nu}(I_\mu + I_\nu)\langle \mu, \boldsymbol{R}_i|\nu, \boldsymbol{R}_j\rangle \; for \; \boldsymbol{R}_i \neq \boldsymbol{R}_j, \tag{S8}$$

$$= I_\mu \delta_{\mu\nu} \quad for \; \boldsymbol{R}_i = \boldsymbol{R}_j. \tag{S9}$$

We use the ionization energies of $I_{3s} = -18.6$ eV and $I_{3p} = -14$ eV, and set $K_{ss} = K_{pp} = 1.75$. As in Ref. 4, we determine $K_{sp} = 1.432$ so that bandgap energy at Z point becomes ~0.3 eV. In the calculation of band structure, we consider up to the intra-layer eighth nearest neighbor and inter-layer third nearest neighbor hopping. The resultant band structure is shown in Figs. S8(a). Since black phosphorus contains 20 valence electrons in its primitive cell, the tenth and eleventh bands respectively correspond to the highest valence (labeled by $v$) and the lowest conduction (labeled by $c$) bands.

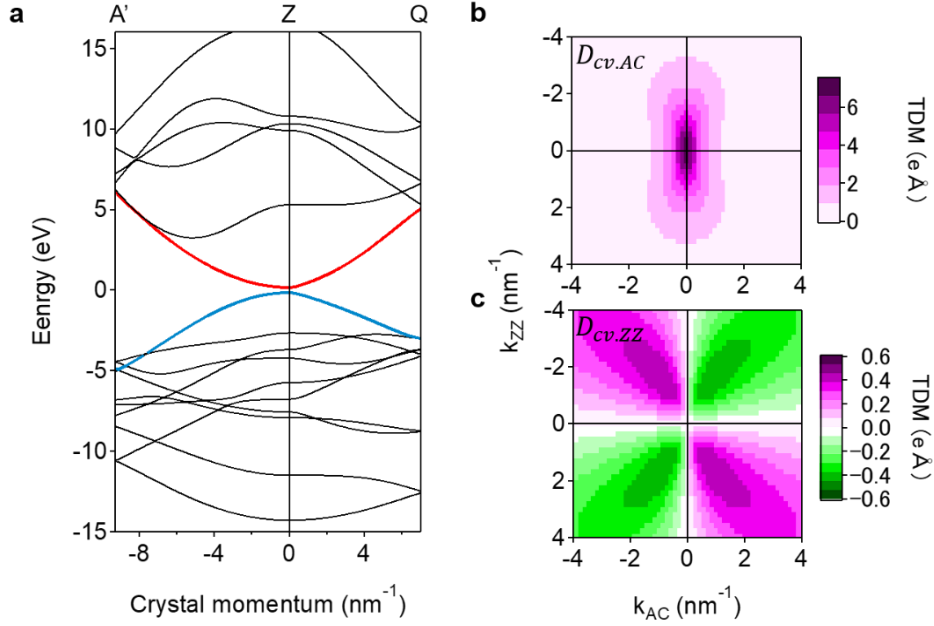

FIG S8: (a) The calculated band structure based on tight-binding method as a function of crystal momentum along Z-Q (>0) and Z-A'(<0) directions. The origin is set to be Z point. Blue solid line indicates highest valence band, and red solid line indicates the lowest conduction band. (b) Two-dimensional map of transition dipole moment along AC direction in momentum space. (c) Two-dimensional map of transition dipole moment along ZZ direction in momentum space. In (b) and (c), the origin is set to be Z point.

Then, we used the eigenfunctions of the highest valence and lowest conduction bands ($|v, \bm{k}\rangle$, $|c, \bm{k}\rangle$) obtained by solving Eq. (S2) for the evaluation of the TDM as follows:

$$\bm{D}_{cv}(\bm{k}) = -\frac{ie\hbar \langle c, \bm{k}|\hat{\bm{p}}|v, \bm{k}\rangle}{m_0(\varepsilon_c(\bm{k}) - \varepsilon_v(\bm{k}))} = \frac{e\hbar^2 \langle c, \bm{k}|\nabla|v, \bm{k}\rangle}{m_0 \varepsilon_g(\bm{k})},$$

(S11)

$$\langle c, \bm{k}|\nabla|v, \bm{k}\rangle = \sum_{i,j,\mu,\nu} C_{i,\mu}^{c*}(\bm{k}) C_{j,\nu}^{v}(\bm{k}) \sum_{\Delta \bm{R}_{ij}} e^{i\bm{k}\cdot \Delta \bm{R}_{ij}} \langle \mu, \bm{0}|\nabla|\nu, \Delta \bm{R}_{ij}\rangle.$$

(S12)

Here, we only consider the first nearest neighbor hopping for the calculation of the TDM for simplicity. Figures S8(b) and (c) respectively show two-dimensional maps of AC and ZZ components of the TDM in crystal momentum given by $\boldsymbol{k} = (k_{AC}, k_{ZZ}, 2\pi/c)$, where c is the layer spacing of bulk BP. The corresponding vector plot is shown in Fig. 4(c) in the main text. Note that the TDM has a complex value in general, and any phase in complex space can be allowed. Here, we choose the phase so that AC component of TDM has a positive real number. Due to the inversion symmetry of the crystal, the ZZ component of the TDM also becomes a real number. In this case, in the region of quadrant I in momentum space represented by $(+, +)$, the sign of the TDM is given by $(+, -)$ as shown in Figs. S8 (b) and (c). This texture determines the important feature of the HHG polarization in BP (Fig. 3 (d) in the main text), and reflects the inter-atomic bonding between the first nearest-neighbor atoms.

7. **Numerical calculation of HHG signal**

We calculate the current in bulk BP under the MIR driving field. For simplicity, we only pick up the highest valence and the lowest conduction bands in BP, and consider temporal evolution of the density matrix of the two-band system with the inversion symmetry as follows:

$$\hbar \frac{\partial}{\partial t} \rho_{vv}(\boldsymbol{k}, t) = e\boldsymbol{F}(t) \cdot \frac{\partial}{\partial \boldsymbol{k}} \rho_{vv}(\boldsymbol{k}, t) + 2Im[\boldsymbol{D}^*_{cv}(\boldsymbol{k}) \cdot \boldsymbol{F}(t) \rho_{cv}(\boldsymbol{k}, t)],$$

(S13)

$$\hbar \frac{\partial}{\partial t} \rho_{cc}(\boldsymbol{k}, t) = e\boldsymbol{F}(t) \cdot \frac{\partial}{\partial \boldsymbol{k}} \rho_{cc}(\boldsymbol{k}, t) - 2Im[\boldsymbol{D}^*_{cv}(\boldsymbol{k}) \cdot \boldsymbol{F}(t) \rho_{cv}(\boldsymbol{k}, t)],$$

(S14)

$$\hbar \frac{\partial}{\partial t} \rho_{cv}(\boldsymbol{k}, t) = -i(\varepsilon_g(\boldsymbol{k}) + ie\boldsymbol{F}(t) \cdot \frac{\partial}{\partial \boldsymbol{k}} - i\gamma) \rho_{cv}(\boldsymbol{k}, t)$$

$$-i\boldsymbol{D}_{cv}(\boldsymbol{k}) \cdot \boldsymbol{F}(t)(\rho_{vv}(\boldsymbol{k}, t) - \rho_{cc}(\boldsymbol{k}, t)).$$

(S15)

Here, we introduce light-matter interaction Hamiltonian by using the dipole approximation with the length gauge $\hat{H}_I(t) = e\hat{r} \cdot F(t)$, where $F(t)$ is the temporal profile of MIR electric field. $\gamma$ is the empirical damping constant, which describes decoherence of the interband polarization. Then, these equations can be transformed into equations given by

$$\hbar \frac{d}{dt}\tilde{\rho}_{vv}(\mathbf{k}_0, t) = 2Im[\mathbf{D}^*_{cv}(\mathbf{K}(t)) \cdot \mathbf{F}(t)\tilde{\rho}_{cv}(\mathbf{k}_0, t)], \tag{S16}$$

$$\hbar \frac{d}{dt}\tilde{\rho}_{cc}(\mathbf{k}_0, t) = -\hbar \frac{d}{dt}\tilde{\rho}_{vv}(\mathbf{k}_0, t), \tag{S17}$$

$$\hbar \frac{d}{dt}\tilde{\rho}_{cv}(\mathbf{k}_0, t) = -i\big(\varepsilon_g(\mathbf{K}(t)) - i\gamma\big)\tilde{\rho}_{cv}(\mathbf{k}_0, t),$$

$$-i\mathbf{D}_{cv}(\mathbf{K}(t)) \cdot \mathbf{F}(t)(\tilde{\rho}_{vv}(\mathbf{k}_0, t) - \tilde{\rho}_{cc}(\mathbf{k}_0, t)), \tag{S18}$$

$$\tilde{\rho}_{ij}(\mathbf{k}_0, t) = \rho_{ij}(\mathbf{K}(t), t), \tag{S19}$$

$$\mathbf{A}(t) = -\int^t dt' \, \mathbf{F}(t'), \tag{S20}$$

$$\mathbf{K}(t) = \mathbf{k}_0 + \frac{e}{\hbar}\mathbf{A}(t). \tag{S21}$$

We define the current operator $\hat{\mathbf{j}}(t)$ as

$$\hat{\mathbf{j}}(t) = -\frac{e\hat{\mathbf{p}}_{kin}}{m} = ie\frac{[\hat{\mathbf{r}}, \hat{H}(t)]}{\hbar}. \tag{S22}$$

Then, the total current induced by the MIR electric field can be written by [6]

$$\mathbf{J}(t) = Tr[\hat{\rho}(t)\hat{\mathbf{j}}(t)] = \sum_k Tr[\rho(\mathbf{k}, t)j(\mathbf{k}, t)]$$

$$= \sum_{\mathbf{k}_0} Tr[\tilde{\rho}(\mathbf{k}_0, t)\tilde{j}(\mathbf{k}_0, t)]. \tag{S23}$$

The total current can be described as the sum of the intraband current $J_{ra}(t)$ and the interband current $J_{er}(t)$ as follows:

$$J_{ra}(t) = -e \sum_{k_0, n=c,v} v_n(K(t)) \tilde{\rho}_{nn}(k_0, t), \quad (S24)$$

$$J_{er}(t) = -\frac{2}{\hbar} \sum_{k_0} \varepsilon_g(K(t)) Im[D_{cv}(K(t))\tilde{\rho}_{cv}^*(k_0, t)], \quad (S25)$$

$$v_n(k) = \frac{\partial}{\hbar \partial k'} \varepsilon_n(k') \bigg|_{k'=k}. \quad (S26)$$

Here, $v_n(k)$ is the group velocity of Bloch electron in $n$th band. We assume that driving MIR electric field is plane wave, therefore, HHG spectrum can be calculated by using following equation:

$$I_{HHG}(\omega) \propto \left| \int dt\, e^{i\omega t} J(t) \right|^2. \quad (S27)$$

We simulated by using the in-plane MIR electric field which has a Gaussian temporal envelope with a full width at half maxima of 60 fs, which corresponds to that of the experiment. The central photon energy of MIR pulse is set to be 0.26 eV, which is same condition with that of our experiment. Since the empirical damping constant $\gamma$ cannot be determined through experiment, we set $\gamma = 0.07$ eV so that spectral modification due to the interference of trajectories becomes negligible. For simplicity, we only consider two-dimensional in-plane momentum space which is represented by $k = (k_{AC}, k_{ZZ}, 2\pi/c)$. To avoid the complicated band crossing point, we set the electric field strength and the region of calculation as follows: $F_0 = 4.5$ MV/cm, $-0.7\pi/l_x < k_{AC} < 0.7\pi/l_x$ and $-0.7\pi/l_y < k_{ZZ} < 0.7\pi/l_y$. Here, $l_x = 2(a_1 \cos \alpha/2 + a_2 \cos \beta)$ and $l_y = 2a_1 \sin \alpha/2$ (see Fig. S7). We used $61 \times 61$ k-point mesh in the simulation results shown below.

Figure S9(a) shows the typical HHG spectrum obtained by numerically solving the temporal evolution of the two-band system for $\theta = 35°$. The spectrum has a qualitative good agreement with experiment: a gradual decrease of HHG yields with the increase of the harmonic order. In addition, there are slight blue shifts of peak energy for the higher harmonics, which are also observed in our experiment (Fig. 1(d) in the main text). This may be caused by the accumulated electron-hole dynamics under temporally periodic driving, which violates the periodicity of emissions [7]. In addition, we compare the contribution of the inteband polarization with that of the intraband current for MIR polarization along AC direction ($\theta = 0°$) as shown in Fig. S9(b). The interband polarization contribution is greater than that of the intraband current especially for the higher harmonics.

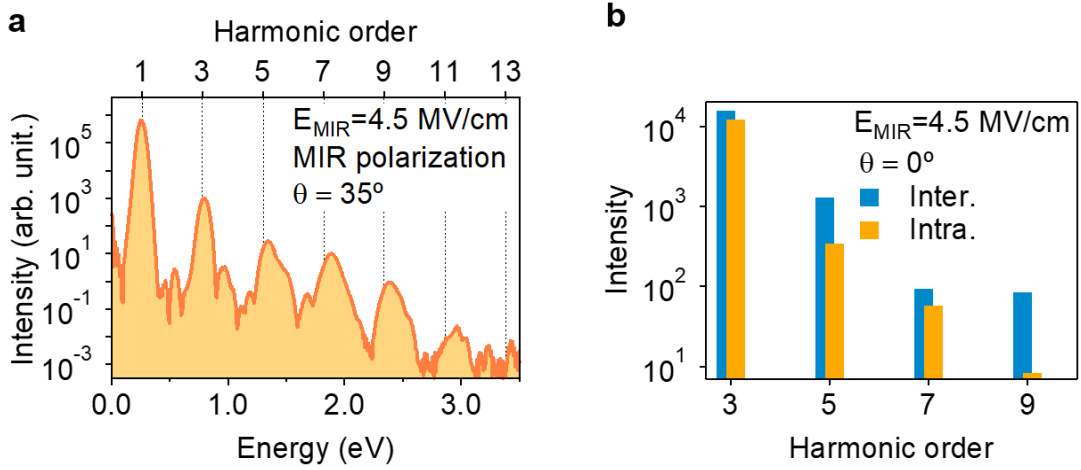

FIG S9: (a) The calculated HHG spectrum with MIR polarization angle $\theta = 35°$ and MIR electric field strength of 4.5 MV/cm. (b) Comparison of the HHG intensities of intrerband polarization (blue) and intraband current (yellow) contributions. MIR polarization is along AC direction $\theta = 0°$, and MIR electric field strength is 4.5 MV/cm.

Figures S10(a-d) respectively show the crystal orientation dependences of HHG yields for the third, fifth, seventh, and ninth harmonics by using the band structure and the TDM texture calculated through the tight-binding model discussed above. In the all harmonics, the contributions of interband polarization are greater than those of intraband current, suggesting the validity of interband resonant HHG regime. For the MIR field along ZZ direction, the HHG yields show strong suppression, and these are good agreement with those of the experiment. However, the suppression of HHG

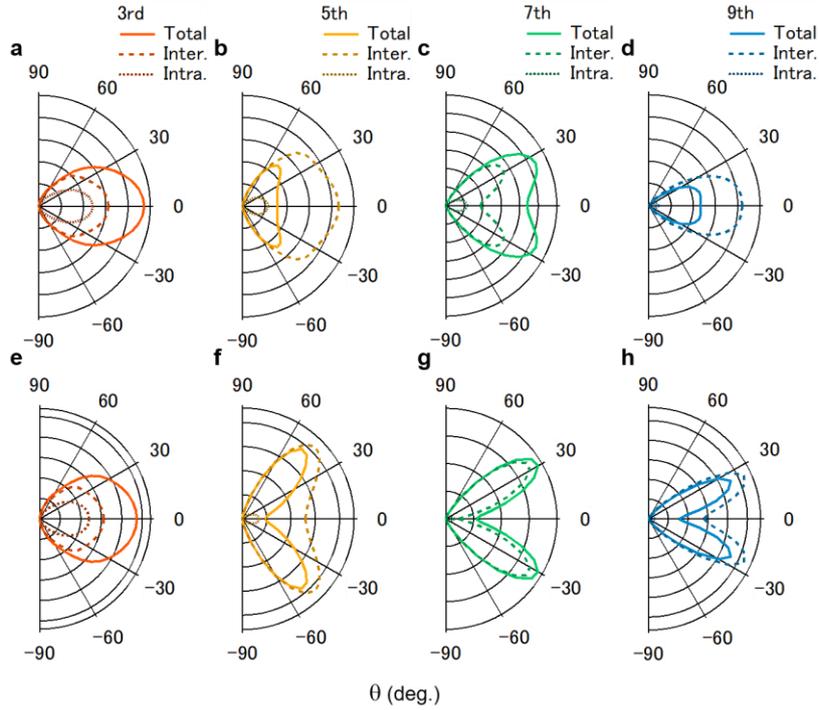

FIG S10: (a-d) The simulated crystal orientation dependence of (a) the third, (b) fifth, (c) seventh, and (d) ninth harmonics with band structure and TDM texture calculated by using tight-binding model. (e-f) The corresponding ones of (e) the third, (f) fifth, (g) seventh, and (h) ninth harmonics with three times larger ZZ component of TDM than that in (a-d). Solid lines indicate the results of total current. Dashed and dotted lines respectively indicate the contribution of interband polarization and intraband current. All the results are calculated by using MIR electric field strength of 4.5 MV/cm.

yields for MIR field along AC direction, which is observed in our experiment for the fifth and higher harmonics, is much weaker in the calculation. These results imply that the calculated TDM texture is deviated from those in the real system.

Figures S10(e-h) respectively show the same crystal orientation dependences with the three times larger ZZ component of the TDM texture (see Fig. S8(c) in Sec. 6 of the Supplemental Information). The fifth and higher harmonics yields show clear suppression for the MIR field along AC direction, which is a qualitative agreement with the experimental results. This is because ZZ

component of the TDM is zero for $k_{ZZ} = 0$ as discussed in Sec. 6. Note that ZZ component of the group velocity of electron-hole, which is directly related to intraband current, also becomes zero for $k_{ZZ} = 0$ (see Sec. 10 for the detail). Therefore, the relative increase of ZZ component of the group velocity to AC component is also expected to cause the suppression of HHG yield for $\theta = 0°$.

The polarization states of HHG help us to distinguish these contributions in HHG process. Figures S11(a) and (b) respectively show the polarization states of the third and fifth harmonics for the MIR polarization along AC direction ($\theta = 0°$). Regardless of the mechanism and the detail of the band structure and the TDM texture, HHG emissions are polarized along AC direction. This polarization state is determined by the symmetry of the system, which is discussed in Sec. 11. On the other hand, for the MIR polarization angle of $\theta = 40°$, the system under the MIR electric field does not hold any symmetry except for trivial one. Therefore, the HHG polarizations for $\theta = 40°$ should highly depend on the detail of the microscopic HHG process.

Figures S11(c) and (d) respectively show the polarization states of the third and fifth harmonics for $\theta = 40°$. The third harmonics are almost polarized along AC direction for both interband polarization and intraband current, but there is slight difference between them. The long axis of interband polarization points the region of quadrant IV (II). In contrast, that of intraband current points quadrant I (III). This tendency is more pronounced in the fifth harmonics (Fig. S11(d)), and the contribution of interband polarization qualitatively corresponds to the experimental one.

Figure S11(e) and (f) respectively show the corresponding results to (c) and (d) calculated with the three times larger ZZ component of TDM. In addition to its HHG yield, the HHG polarization of total current and interband polarization also have a good agreement with the experiment: HHG polarization is almost perpendicular to incident MIR polarization. This also suggests that the contribution of interband polarization dominates HHG process in BP, and the ZZ component of the TDM with respective to the AC component in the real system is several times larger than that of our calculation based on the tight-binding model. The details about the origin of HHG properties are discussed in Secs. 8 and 10.

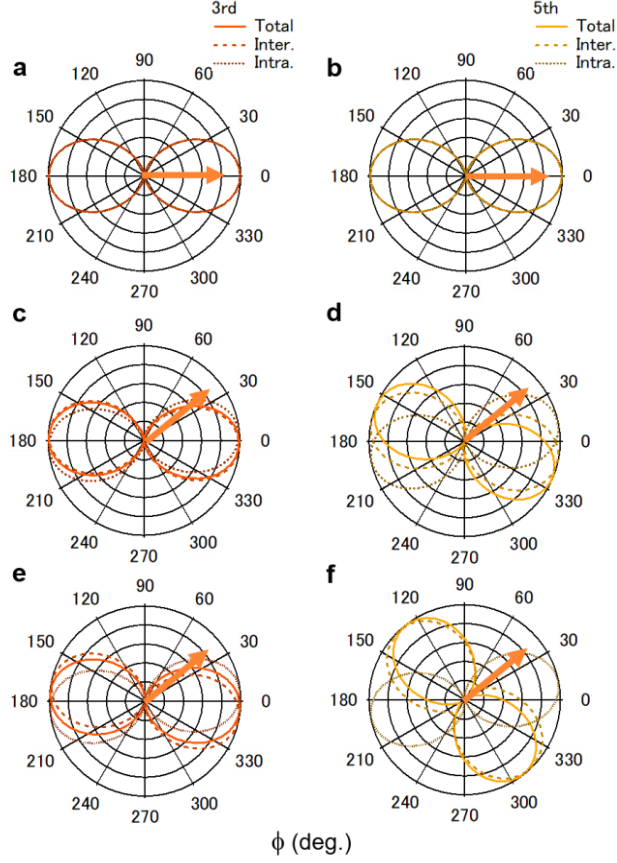

FIG S11: (a,b) Normalized polarization state of (a) the third and (b) fifth harmonics emissions for MIR electric field along AC direction ($\theta = 0°$). (c,d) Normalized polarization state of (c) the third and (d) fifth harmonics for MIR polarization angle $\theta = 40°$ calculated by using band structure and TDM of tight-binding model. (e,f) Normalized polarization state of (e) the third and (f) fifth harmonics for MIR polarization angle $\theta = 40°$ calculated by using three times larger ZZ component of TDM than that in (c) and (d). Dashed and dotted lines respectively indicate the contribution of interband polarization and intraband current. Solid lines indicate the sum of interband polarization and intraband current. The orange arrows indicate the direction of incident MIR polarization. All the results are calculated by using MIR electric field strength of 4.5 MV/cm.

## 8. Interband resonant HHG in bulk BP

Here, instead of numerically solving the temporal evolution of the density matrix, we evaluate the integral form of the solution of the interband polarization current with several approximations and assumptions [8-10]. Especially, we consider the condition that incident MIR field is resonant with the band gap energy of bulk BP, which we call "interband-resonant" HHG in this paper, and derive Eq. (1) in the main text. The validity of Eq. (1) is confirmed by the comparison between analytical and numerical results.

Firstly, for simplicity, we assume that $\tilde{\rho}_{vv}(\boldsymbol{k}_0, t) - \tilde{\rho}_{cc}(\boldsymbol{k}_0, t)$ is almost unity during MIR electric field driving [9,10]. Note that resonant excitation causes efficient creation of electron-hole pairs, and induces depletion of $\tilde{\rho}_{vv}(\boldsymbol{k}_0, t) - \tilde{\rho}_{cc}(\boldsymbol{k}_0, t)$. We discuss the depletion effect on the analysis later. The integral form of interband spectrum can be written as follows:

$$\boldsymbol{J}_{er}(\omega) = -\frac{iV_{cell}}{\hbar^2(2\pi)^3}\int dt \int_{BZ} d\boldsymbol{k}\, \varepsilon_g(\boldsymbol{k})\boldsymbol{D}^*_{cv}(\boldsymbol{k})$$

$$\times \int_{t_0}^t dt'\, \boldsymbol{D}_{cv}\left(\boldsymbol{k}(t,t')\right)\cdot \boldsymbol{F}(t')e^{-iS(\boldsymbol{k},t,t')+i\omega t} + \text{c.c.,} \qquad (S28)$$

$$S(\boldsymbol{k},t,t') = \int_{t'}^t d\tau\, \varepsilon_g(\boldsymbol{k}(t,\tau))/\hbar, \qquad (S29)$$

$$\boldsymbol{k}(t,t') = \boldsymbol{k} + \frac{e}{\hbar}\boldsymbol{A}(t') - \frac{e}{\hbar}\boldsymbol{A}(t). \qquad (S30)$$

Here, the summation over $\boldsymbol{k}$ is approximated by the integral over $\boldsymbol{k}$ in Eq. (S25) and $S(\boldsymbol{k},t,t')$ is the action. We consider the resonant excitation of the bandgap energy. This condition is that in our experimental setup. Note that photon energy is usually far below transition energy in conventional HHG experiments. We apply so-called rotating wave approximation, which is valid when incident light is resonant with the transition energy, as follows:

$$F(t) = F_0 \cos \Omega t \sim \frac{1}{2} F_0 e^{-i\Omega t}. \tag{S31}$$

This gives the interband current spectrum as follows:

$$J_{er}(\omega) = -i \frac{V_{cell}}{2\hbar^2 (2\pi)^3} \int dt \int_{BZ} d\mathbf{k} \varepsilon_g(\mathbf{k}) \mathbf{D}_{cv}^*(\mathbf{k})$$
$$\times \int_{t_0}^{t} dt' \, \mathbf{D}_{cv}\left(\mathbf{k}(t,t')\right) \cdot \mathbf{F}_0 e^{-iS(\mathbf{k},t,t') + i\omega t - i\Omega t'} + c.c.. \tag{S32}$$

Then, we use stationary phase approximation (SPA) for the evaluation of the integral in Eq. (S32). The saddle point equations are given by

$$\frac{\partial}{\partial \mathbf{k}} S\left(\mathbf{k}, t, t'\right) = \int_{t'}^{t} d\tau (\mathbf{v}_c(\mathbf{k}(t,\tau)) - \mathbf{v}_v(\mathbf{k}(t,\tau))) = 0, \tag{S33}$$

$$\frac{\partial}{\partial t'} S\left(\mathbf{k}, t, t'\right) = -\frac{\varepsilon_g\left(\mathbf{k}(t,t')\right)}{\hbar} = -\Omega, \tag{S34}$$

$$\frac{\partial}{\partial t} S\left(\mathbf{k}, t, t'\right) = \frac{\varepsilon_g(\mathbf{k})}{\hbar} - \frac{e\partial A(t)}{\hbar \partial t} \frac{\partial}{\partial \mathbf{k}} S\left(\mathbf{k}, t, t'\right) = \frac{\varepsilon_g(\mathbf{k})}{\hbar} = \omega. \tag{S35}$$

By using the combination of parameters $s = (\mathbf{k}_{st}, t_r, t_i)$ which satisfy the above saddle point equations, $J_{er}(\omega)$ is approximately given by

$$J_{er}(\omega) \approx -i \frac{V_{cell}}{2\hbar(2\pi)^3} \omega \sum_s \mathbf{D}_{cv}^*(\mathbf{k}_{st}) \mathbf{D}_{cv}(\mathbf{k}_i) \cdot \mathbf{F}_0$$
$$\times \frac{\exp(-iS(\mathbf{k}_{st}, t_r, t_i) + i\omega t_r - i\Omega t_i)}{\sqrt{|\hat{S}''(\mathbf{k}_{st}, t_r, t_i)|}}, \tag{S36}$$

Here, $\hat{S}''(\mathbf{k}_{st}, t_r, t_i)$ is the Hessian matrix of $S\left(\mathbf{k}, t, t'\right)$. The important difference from conventional HHG process is represented in Eq. (S34). In conventional HHG process, right hand side of Eq. (S34) is set to be zero, and only complex-valued solutions are allowed. The complex-valued solution corresponds to the quantum-tunneling process for electron-hole creation. In contrast, since $\hbar\Omega$ is equal to bandgap energy at the band edge $\varepsilon_g(\mathbf{k}_i)$ in our condition, the equation has a real-valued solution as follows:

$$\mathbf{k}_{st} = \mathbf{k}_i - \frac{e}{\hbar} \mathbf{A}(t') + \frac{e}{\hbar} \mathbf{A}(t). \tag{S37}$$

This reflects that electron-hole pairs are created at band edge $\mathbf{k}_i = (0,0,2\pi/c)$ without tunneling process, and then, are accelerated in the direction that electric field is applied. Equation (S35) represents the conservation of energy in emission process. At recombination time $t_r$, the high harmonics are emitted with the energy equal to the transition energy: $\varepsilon_g(\mathbf{k}_i - e\mathbf{A}(t_i)/\hbar + e\mathbf{A}(t_r)/\hbar)$. In the region where the transition energy $\varepsilon_g(\mathbf{k})$ monotonically increases with the increase of $|\mathbf{k} - \mathbf{k}_i|$, there is only two solutions which satisfy Eq. (S35): $\mathbf{k}_r$ or $-\mathbf{k}_r + 2\mathbf{k}_i$. When MIR electric field is polarized along the in-plane direction, we can set $\mathbf{D}^*_{cv}(-\mathbf{k}_r + 2\mathbf{k}_i) = \mathbf{D}^*_{cv}(\mathbf{k}_r)$ from the symmetry of the crystal in BP. Therefore, there is one-to-one correspondence between the emitted HHG polarization and $\mathbf{D}^*_{cv}(\mathbf{k}_r)$ for the real-valued solution of Eqs. (S34) and (S35) as follows:

$$\mathbf{J}_{er}(\omega) \approx -i \frac{V_{cell}}{2\hbar(2\pi)^3} \omega \mathbf{D}^*_{cv}(\mathbf{k}_r) \mathbf{D}_{cv}(\mathbf{k}_i) \cdot \mathbf{F}_0 \sum_s \frac{\exp(-iS(\mathbf{k}_r,t_r,t_i)+i\omega t_r - i\Omega t_i)}{\sqrt{|\hat{S}''(\mathbf{k}_r,t_r,t_i)|}},$$
(S38)

As discussed in Uzan et al. [8], the determinant of Hessian $|\hat{S}''|$ is represented by the curvature of the energy band at emission point $\mathbf{k}_r$, and causes a strong enhancement of HHG emission where the curvature becomes zero, i.e., van-Hove singular points. In the resonant excitation case, in addition to emission process, creation process is also important since the determinant of Hessian is also represented by the curvature of energy band at creation point $\mathbf{k}_i$. In our experimental condition, the determinant of Hessian become zero for the solutions which satisfy Eq. (S37), i.e., real-valued solutions of Eq. (S34) owing to the zero group velocity at the band edge. This results in divergence of Eq. (S38) and indicates that the contribution of the real-valued solution of Eq. (S34) plays dominant role in resonant-HHG process. To evaluate the integral of Eq. (S32), we apply sequentially SPA on $t'$ and then on $\{\mathbf{k},t\}$, and obtain the following equation instead of Eq. (S38) [10].

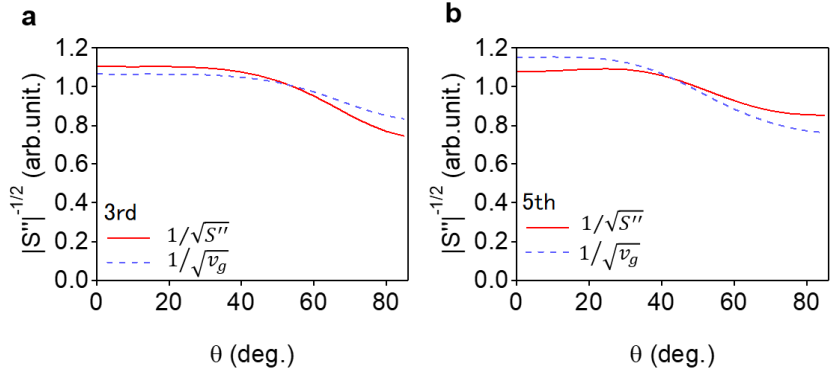

FIG S12: (a,b) Inverse square root of the Hessian $\left|\hat{S}''(k_r, t_r, t_i)_{\{k,t\}}\right|$ for the (a)third and (b)fifth harmonics as a function of MIR field direction evaluated by numerical calculation with MIR electric field of 4.5 MV/cm. Dashed lines indicate the inverse square root of the group velocity at the emission energy. For clarity, all the plots are magnified so that the average values become one.

$$J_{er}(\omega) \approx -i \frac{V_{cell}}{2\hbar(2\pi)^3} \omega D^*_{cv}(k_r) D_{cv}(k_i) \cdot F_0 \sum_s \frac{\exp(-iS(k_r,t_r,t_i)+i\omega t_r - i\Omega t_i)}{\sqrt{\epsilon + \left|\hat{S}''_{\{t'\}}\right|}\sqrt{\left|\hat{S}''_{\{k,t\}}\right|}} .$$

(S39)

Here, we introduce a small regulation constant $\epsilon$ to avoid the divergence, and $\hat{S}''(k_r, t_r, t_i)_{\{t'\}(\{k,t\})}$ is the Hessian with respective to $\{t'\}(\{k,t\})$. For the real-valued solution of Eq. (S34), $\hat{S}''(k_r, t_r, t_i)_{\{t'\}}$ becomes zero.
Then, let us discuss the detail of Eq. (S33), which describes the recombination condition of electron-hole pairs in real space with classical picture. In solid system, $\partial_k \varepsilon_g(k)$ is not parallel to $k$, i.e., two-dimensional motion in real space, and it is difficult to find the real-valued solutions which satisfy all the saddle point equations. However, in our condition, electron-hole pairs are created only at $k_i$ and accelerated along electric field direction. This indicates that electron-holes in real space are delocalized along the perpendicular direction of electric field. This implies that recombination is allowed although classical trajectories of electron and hole do not have intersection. Hence, we modify the restriction of Eq. (S33) to Eq. (S38) for the evaluation of the integral as follows:

$$\int_{t'}^{t} d\tau (v_{c,\parallel}(\boldsymbol{k}(t,\tau)) - v_{v,\parallel}(\boldsymbol{k}(t,\tau))) = 0 \ . \tag{S40}$$

Here, $v_{c,\parallel} - v_{v,\parallel}$ is the group velocity of the relative motion of eletron-hole pair parallel to the field direction. Note that this treatment is approximate one, and the crystal orientation dependence of HHG obtained from Eq. (S38) may be modified by the recombination condition described in Eq. (S33). As a result, interband current can be approximately evaluated as follows:

$$\boldsymbol{J}_{er}(\omega) \approx -i \frac{V_{cell}}{2\hbar(2\pi)^3} \alpha(\theta, F_0, \omega, \Omega) \omega \boldsymbol{D}_{cv}^*(\boldsymbol{k}_r) \boldsymbol{D}_{cv}(\boldsymbol{k}_i) \cdot \boldsymbol{F}_0, \tag{S41}$$

$$\alpha(\theta, F_0, \omega, \Omega) = \sum_s \frac{\exp(-iS(k_r, t_r, t_i) + i\omega t_r - i\Omega t_i)}{\sqrt{\epsilon |\hat{S}''(k_r, t_r, t_i)_{\{k,t\}}|}} \ . \tag{S42}$$

Here, the term of $\boldsymbol{D}_{cv}(\boldsymbol{k}_i) \cdot \boldsymbol{F}_0$ describes electron-hole creation process, $\alpha$ describes electron-hole recombination dynamics, and $\boldsymbol{D}_{cv}^*(\boldsymbol{k}_r)$ describes emission process. As discussed in Uzan et al. [8], the factor $\alpha$ is almost proportional to the inverse square root of group velocity at $\boldsymbol{k}_r$. Figures S12(a) and (b) respectively show the $\alpha(\theta)$ of the third and fifth harmonics calculated by using Eq. (S42) for the short trajectory contribution with the calculated lowest conduction and highest valence bands in Fig. S8(a). The angle dependence is almost same as that of group velocity at emission point, and its anisotropy is weak compared with observed HHG anisotropy. Hence, in the following discussion, we assume that $\alpha$ is constant with respect to $\theta$. Note that this contribution should be taken account for the system with complicated band structure, which has several van Hove singularity points [8].

Figures S13(a-d) respectively show the crystal orientation dependence of (a) the third, (b) fifth, (c) seventh, and (d) ninth harmonics estimated by using Eq. (S41). Solid lines are the results with the tight-binding model, and dotted lines are those with the three times larger ZZ component of the TDM. Reflecting the increase of the ZZ component of TDM texture, the suppression of fifth and higher harmonics yields becomes strong for $\theta = 0°$, which has a good agreement with the experiment.

Equation (S41) gives us an intuitive understanding of these crystal orientation dependence. Since TDM at band edge $\boldsymbol{D}_{cv}(\boldsymbol{k}_i)$ is parallel to AC direction, creation of electron-hole pairs is forbidden for $\theta = 90°$. This causes strong suppression of HHG yields for $\theta = 90°$. As shown in Fig. S8 (b) in Sec. 6, AC component of TDM shows the maximum at Z point and sudden decrease with the increase of the absolute value of $k_{AC}$. Whereas, the ZZ component of TDM has the maximum value for finite $k_{AC}$ and $k_{ZZ}$ (see Fig. S8(c) in Sec. 6 of the Supplemental Information). Therefore, at the crystal momentum far away from Z point, ZZ component of the TDM texture become relatively important compare with that of AC component. Since the ZZ component of the TDM is zero for $k_{ZZ} = 0$, the higher harmonics, where electron-hole pairs far away from Z point involves, show suppression for $\theta = 0°$ in emission process.

Then, we address the effect of the depletion of electrons in valence band $(\tilde{\rho}_{vv}(\boldsymbol{k}_0, t) - \tilde{\rho}_{cc}(\boldsymbol{k}_0, t))$. Since the resonant excitation of the band edge with strong laser field efficiently creates coherent electron-hole pairs, electrons in valence bands gradually decrease under laser irradiation in our experimental condition. Then, the interband current is proportional to $\Delta\tilde{\rho}(\boldsymbol{k}_0, t) = \tilde{\rho}_{vv}(\boldsymbol{k}_0, t) - \tilde{\rho}_{cc}(\boldsymbol{k}_0, t)$ as follows:

$$\boldsymbol{J}_{er}(\omega) = -i\frac{V_{cell}}{2\hbar^2(2\pi)^3}\int dt \int_{BZ} d\boldsymbol{k}\varepsilon_g(\boldsymbol{k})\boldsymbol{D}^*_{cv}(\boldsymbol{k})\int_{t_0}^{t} dt'$$
$$\times \boldsymbol{D}_{cv}\left(\boldsymbol{k}(t,t')\right) \cdot \boldsymbol{F}_0\Delta\tilde{\rho}(\boldsymbol{k} - e\boldsymbol{A}(t)/\hbar, t')e^{-iS(\boldsymbol{k},t,t')+i\omega t - i\Omega t'} + c.c..$$
(S43)

We can apply SPA for slowly varying $\Delta\tilde{\rho}(\boldsymbol{k} - e\boldsymbol{A}(t)/\hbar, t')$, and get an additional factor $\beta(\boldsymbol{k}_r, t_r, t_i)$ in Eq. (S41), which gradually decreases with the temporal translation of $t_i(t_r) \to t_i(t_r) + \pi/\Omega$. This additional factor $\beta$ affects inter-cycle interference, and causes spectral modification, broadening, and the suppression of HHG yields. Therefore, we cannot evaluate absolute value of TDM through the HHG method, but the relative relation between HHG properties and TDM is still valid. Here, we assume that the depletion effect is not sensitive with respect to MIR polarization angle $\beta$.

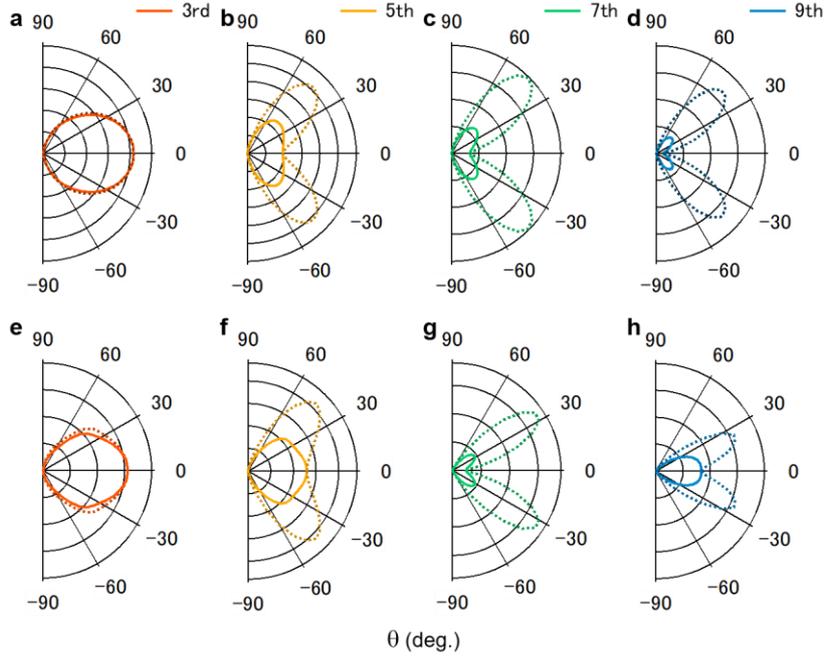

FIG S13: (a-d) The crystal orientation dependence of (a) the third, (b) fifth, (c) seventh, and (d) ninth harmonics evaluated by using Eq. (B12). (e-f) The corresponding interband polarization component of (e) the third, (f) fifth, (g) seventh, and (h) ninth harmonics obtained by solving temporal evolution of two-band system with MIR electric field strength of 4.5 MV/cm. Solid lines indicate the results with TDM calculated by tight-binding model. Dotted line indicate those with three times larger ZZ component of TDM.

Finally, we check the validity of Eq. (S41) thorough the comparison between numerical and analytical calculation. Figures S13(e-h) respectively show the corresponding results to (a-d) calculated by numerically solving the temporal evolution of two band system, which exactly takes account the effect of $\alpha$ and $\beta$. The fifth and higher harmonics yields show suppression with the increase of ZZ component of TDM, which are similar to those obtained by Eq. (S41). This qualitative agreement between numerical simulation and analytical expression suggests the validity of interband-resonant HHG process discussed above and we can reconstruct an approximate TDM texture in two-dimensional Brillouin zone.

## 9. Reconstruction of TDM texture

Let us consider the experimental procedure of the reconstruction of the TDM texture based on Eq. (S41). Here, we use linearly polarized MIR electric field, whose polarization angle is set be $\theta$. Also, the polarization of HHG is resolved by polarizer whose transmission axis angle is set to be $\phi$. The TDM texture is defined as $\boldsymbol{D}_{cv}(\boldsymbol{k}) = D_{cv}(\boldsymbol{k})(\cos\psi(\boldsymbol{k}), \sin\psi(\boldsymbol{k}))$, where $D_{cv}$ and $\psi$ respectively represent the amplitude and the direction of the TDM. Then, HHG intensity $I_{\text{HHG}}(\theta, \phi)$ is given by

$$I_{\text{HHG}}(\theta,\phi) \propto D_{cv}^2(k_r \boldsymbol{e}_\theta)\cos^2(\psi(k_r \boldsymbol{e}_\theta) - \phi) D_{cv}^2(\boldsymbol{k}_i)\cos^2(\psi(\boldsymbol{k}_i) - \theta) \quad , \tag{S44}$$

where $\boldsymbol{e}_\theta = (\cos\theta, \sin\theta)$, and $k_r$ satisfies $\varepsilon_g(k_r \boldsymbol{e}_\theta) = \hbar\omega$. Here, we assume that $\alpha$ in Eq. (S41) is constant with respective to $\theta$. In BP, the TDM at the band edge is parallel to AC direction, i.e., $\psi(\boldsymbol{k}_i) = 0$. This gives

$$\frac{I_{\text{HHG}}(\theta,\phi)}{\cos^2\theta} \propto D_{cv}^2(k_r \boldsymbol{e}_\theta)\cos^2(\psi(k_r \boldsymbol{e}_\theta) - \phi). \tag{S45}$$

Here, by setting $\phi = 0°$ and $90°$ we could obtain TDM texture as follows:

$$D_{cv}(k_r \boldsymbol{e}_\theta) \propto \sqrt{\frac{I_{\text{HHG}}(\theta, 0°) + I_{\text{HHG}}(\theta, 90°)}{\cos^2\theta}}, \tag{S46}$$

$$\psi(k_r \boldsymbol{e}_\theta) = \operatorname{atan}\sqrt{s\frac{I_{\text{HHG}}(\theta, 90°)}{I_{\text{HHG}}(\omega, \theta, 0°)}} \quad s = \pm 1. \tag{S47}$$

We determine sign of $s$ by using the results in Figs. 3 in the main text. Note that we cannot evaluate the absolute value of TDM, but can evaluate relative amplitude and direction of TMD among the same isoenergy line.
As a result, by measuring the crystal orientation dependence of HHG yields and polarizations, we can fully reconstruct $\boldsymbol{D}_{cv}(\boldsymbol{k})$ on isoenergy lines as shown in Fig. S14(a) (Fig. 4(a) in the main text). Compared with the calculation result based on tight binding model (Fig. S14(b)), experimentally obtained one is more tilted towards ZZ direction than the calculated one, i.e., several times relatively larger ZZ component. Figure S14(c) shows corrected TDM texture in which ZZ component is three times multiplied from that of the tight-binding model. The texture is in good agreement with the experimental one.

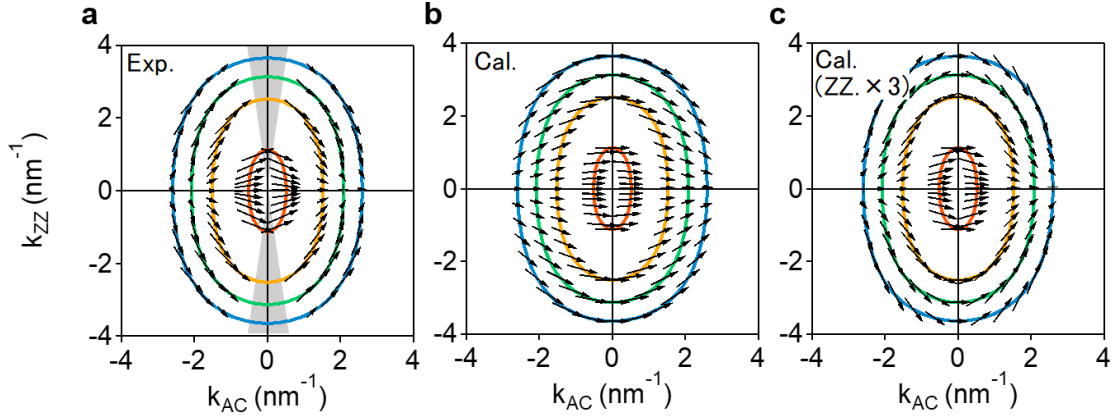

FIG S14: (a) Experimentally obtained TDM texture. The origin is Z point indicated by mirror index of (0, 0, 2π/c). Grey shaded area indicates the region where we cannot obtain TDM due to the small signal to noise ratio. (b) Calculated TDM texture on isoenergy lines corresponding to Figs. S8(b) and (c) in Sec. 6 of the Supplemental Information. (c) Corrected TDM texture in which the ZZ component of TDM is three times multiplied from that in (b). Red, orange, green, and blue solid lines show isoenergy lines corresponding to the emission energy of the $3^{rd}$ (0.8 eV), $5^{th}$ (1.3 eV), $7^{th}$ (1.8 eV), and $9^{th}$ (2.3 eV) harmonics, respectively. Peak amplitude of TDM vector on each isoenergy line is normalized for clarity.

## 10. Intraband current in Bulk BP

Here, we evaluate the intraband current induced by MIR electric field in bulk BP, which has a strong in-plane anisotropy of effective mass. For simplicity, we assume that electron-hole pairs are created at Z point at the initial time, and neglect the temporal evolution of $\tilde{\rho}_{cc}(\mathbf{K},t)$ and $\tilde{\rho}_{vv}(\mathbf{K},t)$. Then, the intraband current in Eq. (S24) can be written by

$$\mathbf{J}_{ra}(t) = -e\mathbf{v}_g\left(\mathbf{K}_Z - \frac{e}{\hbar}\mathbf{A}(t)\right)\tilde{\rho}_{cc}(\mathbf{K}_Z, 0), \tag{S48}$$

$$\mathbf{v}_g(\mathbf{k}) = \mathbf{v}_c(\mathbf{k}) - \mathbf{v}_v(\mathbf{k}), \tag{S49}$$

where $\mathbf{K}_Z$ is the crystal momentum of Z point ($\mathbf{K}_Z = (0,0,2\pi/c)$), $\mathbf{v}_g(\mathbf{k})$ is the group velocity of the relative motion of electron-hole pairs. Figure S15 shows the two-dimensional maps of the group velocity $\mathbf{v}_g(\mathbf{k})$ based on the calculated band structure as shown in Fig. S8(a) in Sec. 6 of the Supplemental Information. The texture is qualitatively different from that of the TDM that determines the properties of the interband HHG (Figs. S8(b) and (c)). Let us only consider the sign of in-plane crystal momentum, TDM, and group velocity. At crystal momentum represented by $(+, +)$ (quadrant I), the sign of group velocity and TDM are respectively given by $(+, +)$ (quadrant I) and $(+, -)$ (quadrant IV). This difference between the TDM and the group velocity gives qualitatively different property of HHG polarization.

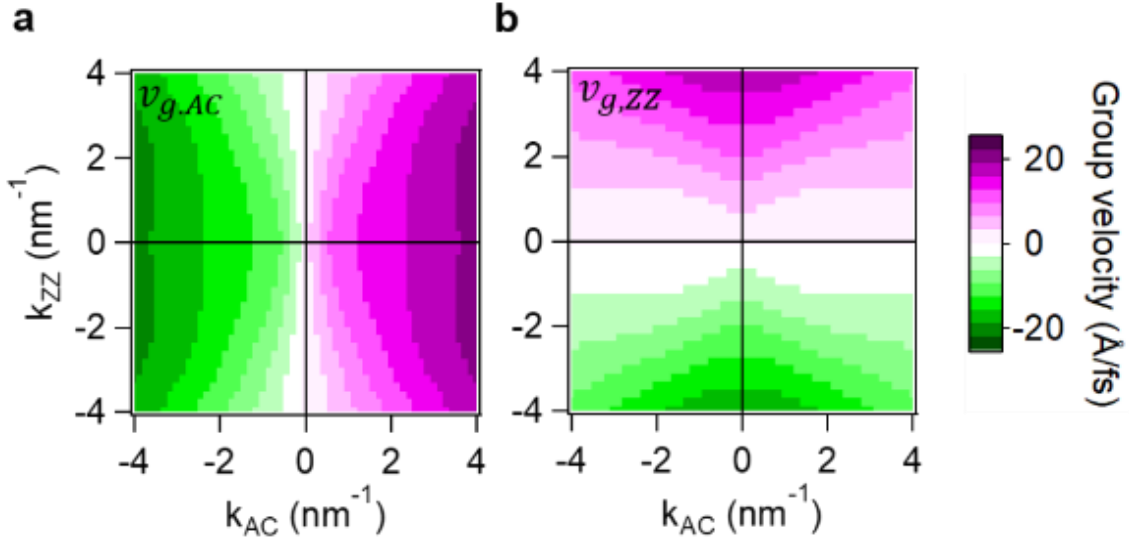

FIG S15: (a) AC component of group velocity of electron-hole pairs as a function of in-plane crystal momentum. (b) ZZ component of group velocity of electron-hole pairs as a function of in-plane crystal momentum. The origin is set to be Z point in both (a) and (b).

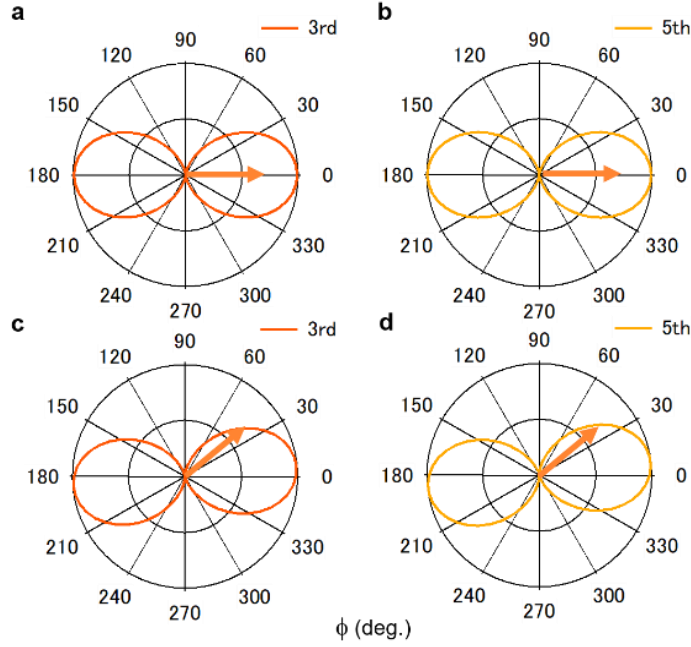

FIG S16: (a,b) Normalized polarization state of (a) the third and (b) fifth harmonics emissions for MIR electric field along AC direction ($\theta = 0°$) by using Eq. (S48). (c,d) Normalized polarization state of (c) the third and (d) fifth harmonics for MIR polarization angle $\theta = 40°$ by using Eq. (S48). All the results are calculated by using MIR electric field strength of 4.5 MV/cm.

Figures S16(a) and (b) respectively show the polarization states of the third and fifth order harmonics calculated by using Eq. (S48) for the incident MIR polarization angle $\theta = 0°$ (AC direction). Both the third and fifth order harmonics are polarized along AC direction, and these results are same as those in the interband HHG. For $\theta = 40°$, both the third and fifth harmonics are also polarized almost along AC direction, but are slightly directed to the region of quadrant I $(+, +)$, which is same as the incident MIR polarization (Figs. S16(c) and (d)). These results reflect the texture of the group velocity in momentum space, and is inconsistent with those of the experiment and the calculated interband HHG as shown in Figs. 3 and in Figs. S11. This strongly support our claim that the contribution of the intraband current to total HHG emission is negligible.

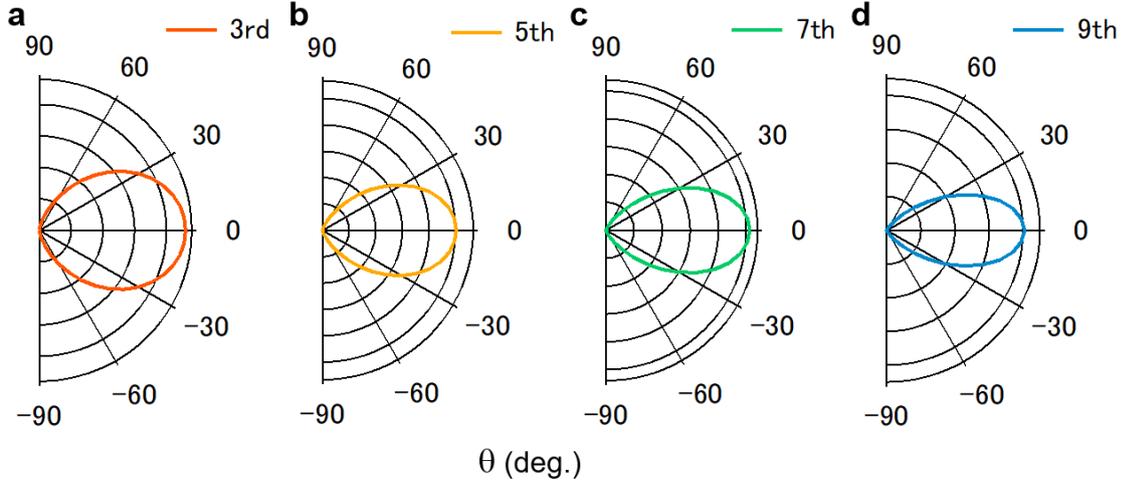

FIG S17: (a-d) The crystal orientation dependence of (a) the third, (b) fifth, (c) seventh, and (d) ninth harmonics evaluated by using Eq. (S48). All the results are calculated by using MIR electric field strength of 4.5 MV/cm.

Figures S17 show the calculated crystal orientation dependence of the HHG yields based on the intraband current model. Almost all the HHG yields show the maximum for the MIR polarization angle $\theta = 0°$. For the third harmonics, this is consistent with the experimental result. However, for the higher harmonics, this result is totally different from the experimental results, which shows strong suppression when MIR polarization is along AC direction ($\theta = 0°$). In addition to HHG polarization, HHG yield depending on the crystal orientation suggests that intraband current is negligible in HHG process.

## 11. Symmetry analysis of polarization selection rule

By considering the dynamical symmetry (DS) of the system, we can derive the polarization selection rule of HHG [11, 12]. For simplicity, we assume that the driving field points in-plane direction, and is a continuous wave, i.e., time-dependent Hamiltonian of the system has a temporal periodicity of $2\pi/\Omega$.

When a so-called Floquet state is realized in the system, the induced current $\boldsymbol{J}(t)$ should uphold the DS as follows:

$$\boldsymbol{J}(t) = \hat{X}\boldsymbol{J}(t). \tag{S50}$$

Here, $\hat{X}$ is a DS operation. Since $\boldsymbol{J}(t)$ also has a temporal periodicity of $2\pi/\Omega$ in the Floquet system, $\boldsymbol{J}(t)$ can be written by

$$\boldsymbol{J}(t) = \sum_{n=1}^{\infty} \sum_{i=AC,ZZ} J_{n,i} \boldsymbol{e}_i \cos(n\Omega t + \phi_{n,i}), \tag{S51}$$

where $n$ is the harmonic order. The polarization selection rule of HHG can be determined by substituting Eq. (S51) for Eq. (S50). Hereafter, we consider several specific DS operations in the system which determines the polarization selection rule of HHG in bulk BP.

Since the crystal symmetry of bulk BP belongs to $D_{2h}$ point group, the system without MIR electric field has an inversion symmetry. Thus, the system under MIR electric field driving has a DS operation as follows:

$$\hat{X}_i = \hat{\tau}_2 \cdot \hat{\iota}. \tag{S52}$$

Where $\hat{\tau}_m$ is temporal translation by $2\pi/m\Omega$ and $\hat{\iota}$ is spatial inversion. This gives the restrictions as follows:

$$J_{n,i} = -J_{n,i} = 0 \text{ for } n\text{: even}. \tag{S53}$$

These indicate that even-order harmonics are forbidden in the crystal with the inversion symmetry.

Next, let us consider that MIR electric field is linearly polarized and can be written by $\mathbf{E}_0 \cos \Omega t$. Since the system has a time-reversal symmetry, the restriction is given by

$$\phi_{n,i} = 0. \tag{S54}$$

This means that HHG is emitted without phase retardation to MIR electric field, and thus, the polarization of HHG is also linearly polarized.

The above discussions can be applied to all experimental results in the main paper. Note that the resonant excitation of the band edge causes a real carrier generation in the system, and slow temporal evolution of the system with respective to temporal period of MIR driving field can occur. Thus, the temporal translation and time-reversal symmetries are not perfectly held in the real system, resulting in slight energy shift of HHG emission, and elliptically polarized emission observed both in experiment and simulation.

Then, let us consider that MIR electric field is polarized along AC direction. The system is identical to that with the mirror operation of $\hat{\sigma}_{AC-Z}$ with the mirror plane that includes AC-axis and out of plane (ZZ) axis. This gives

$$J_{n,ZZ} = -J_{n,ZZ} = 0, \tag{S55}$$

in other word, HHG emission is polarized along AC direction. This result is consistent with those of the experiment and the simulation.

For the MIR electric field along ZZ direction, the system has the DS operation of $\hat{\tau}_2 \cdot \hat{\sigma}_{AC-Z}$. This gives

$$J_{n,AC} = -J_{n,AC} = 0 \text{ for } n\text{: odd}, \tag{S56}$$

in other word, HHG emission is polarized along ZZ direction.

Except for MIR polarization along AC and ZZ direction, there is no restriction to the HHG polarization in addition to Eqs. (S50) and (S51). This indicates that HHG polarization is allowed to point any direction, and this is consistent with our experimental result; almost perpendicular HHG polarization to incident MIR polarization. HHG properties which is not determined by the symmetry of the system gives us hints about the microscopic electron dynamics under strong laser field, and the insight of the internal structure of solid system.